\documentclass[11pt,reqno]{amsart}

\usepackage[utf8]{inputenc}

\usepackage{geometry}
\usepackage{fullpage}
\usepackage{cite}
 
\usepackage{amsmath,amsthm,amssymb}
\usepackage{graphicx}
\usepackage{mathabx}
\usepackage{color}
 
\newcommand{\bPsi}{{\bf\Psi}}

\newcommand{\coup}{{g}} 
\newcommand{\br}{{\bf r}}

\newcommand{\bq}{{\bf q}}

\newcommand{\bv}{{\bf v}}

\newcommand{\bk}{{\bf k}}

\newcommand{\bJ}{{\bf J}}
\newcommand{\bK}{{\bf K}}
\newcommand{\bX}{{\bf X}}

\newcommand{\bx}{{\bf x}}
\newcommand{\by}{{\bf y}}

\newcommand{\hbk}{{\hat{\bk}}}

\newcommand{\dd}{{{d}}}

\newcommand{\im}{\mathrm{Im} \, }
\newcommand{\re}{\mathrm{Re} \, }

\title{Kinetic equations for two-photon light in random media}

\author{Joseph Kraisler}
\address{Department of Applied Physics and Applied Mathematics, Columbia University, New York, NY 10027}
\email{jek2199@columbia.edu}

\author{John C. Schotland}
\address{Department of Mathematics and Department of Physics, Yale University, New Haven, CT 06515}
\email{john.schotland@yale.edu}

\date{\today}
\begin{document}

\begin{abstract}
We consider the propagation of light in a random medium of two-level atoms. We investigate the dynamics of the field and atomic probability amplitudes for a two-photon state and show that at long times and large distances,  the corresponding average probability densities can be determined from the solutions to a system of kinetic equations.

\end{abstract}

\maketitle


\section{Introduction}

The propagation of light in random media is usually considered within the setting of classical optics~\cite{vanRossum_1999}. However, there has been considerable recent interest in phenomena where quantum effects play a key role~\cite{Lodahl_2005_1,Smolka_2009, Smolka_2012,Peeters_2010,Pires_2012,Smolka_2011,vanExter_2012,Lodahl_2005_2,Lodahl_2006_1,Lodahl_2006_2,Patra_1999,Beenakker_2009,Tworzydlo_2002,Beenakker_1998}. Of particular importance is understanding the  impact of multiple scattering on entangled two-photon states, with an eye towards characterizing the transfer of entanglement from the field to matter~\cite{Beenakker_2009,Lahini_2010,Peeters_2010,Ott_2010,Cherroret_2011,Cande_2013,Cande_2014}. Progress in this direction can be expected to lead to advances in spectroscopy~\cite{ Skipetrov_2007}, imaging~\cite{Klyshko_1988,Strekalov_1995,Abouraddy_2001,Abouraddy_2004,Gatti_2004,Scarcelli_2004, Scarcelli_2006,Erkmen_2008, DAngelo_2005,Schotland_2010} and communications~\cite{Moustakas_2000,Skipetrov_2008,Shapiro_2009} .

The propagation of two-photon light is generally investigated either in free space or, in some cases, with account of diffraction~\cite{Saleh_2000,Abouraddy_2002} or scattering~\cite{Schotland_2016}. In this paper, we consider the propagation of two-photon light in a random medium. A step in this direction was taken in~\cite{Markel_2014}, where 
a model in which the field is quantized and the medium is treated classically was investigated. The main drawback of that work is that it is does not allow for the transfer of entanglement between the field and the atoms or between the atoms themselves. Instead, we treat the problem from first principles, employing a model in which the field and the matter are both quantized. We show that for a medium consisting of two-level atoms, the field and atomic probability amplitudes for a two-photon state obey a system of nonlocal partial differential equations with random coefficients. Using this result, we find that at long times and large distances, the corresponding average probability densities in a random medium can be determined from the solutions to a system of kinetic equations. These equations follow from the multiscale asymptotics of the average Wigner transform of the amplitudes in a suitable high-frequency limit~\cite{Ryzhik_1996,Bal_2005,Caze_2015,Carminati_2020}. This formulation of the problem builds on earlier research by the authors on collective emission of single photons in a random medium of two-level atoms~\cite{Kraisler_2022}. In that work, we employed a formulation of quantum electrodynamics in which the field is quantized in real space, thus allowing for the field and atomic degrees of freedom to be treated on the same footing.

This paper is organized as follows. In section 2 we formulate our model for the propagation of two-photon light and derive the equations governing the dynamics of the field and atomic degrees of freedom. Section 3 is concerned with the application of the real-space formalism to the problem of stimulated emission by a single atom. In section 4 we study the dynamics of a two-photon state in a homogeneous medium. In section 5 we discuss the problem of stimulated emission in a random medium and obtain the governing kinetic equations. Section 6 takes up the general problem of two-photon transport in random media and presents the relevant kinetic equations. Our conclusions are formulated in section 7. The appendices contain the details of calculations.

\section{Model}
We consider the following model for the interaction between a quantized massless scalar  field and a system of two-level atoms. The atoms are taken to be identical, stationary and sufficiently well separated that interatomic interactions can be neglected. 
The overall system is described by the Hamiltonian
\begin{equation}
\label{Htot}
H=H_F + H_A + H_I \ , 
\end{equation}
where $H_F$ is the Hamiltonian of the field, $H_A$ is the Hamiltonian of the atoms and $H_I$ is the interaction Hamiltonian.
In order to treat the atoms and the field on the same footing, it is useful to introduce a real-space representation of $H$~\cite{Kraisler_2022}.
The Hamiltonian of the field is of the form 
\begin{align}
\label{H_F}
 H_F = \hbar c\int d^3x (-\Delta)^{1/2}\phi^{\dagger}(\bx)\phi(\bx) \ ,
\end{align}
where $\phi$ is a Bose scalar field that obeys the commutation relations
\begin{align}
\label{commutation}
[\phi(\bx),\phi^{\dagger}(\bx')]&=\delta(\bx-\bx') \ , \\
[\phi(\bx),\phi(\bx')]&=0 \ .
\end{align}
where we have neglected the zero-point energy. The nonlocal operator $(-\Delta)^{1/2}$ is defined by the Fourier integral
\begin{align}
(-\Delta)^{1/2}f(\bx) &= \int\frac{d^3 k}{(2\pi)^3} e^{i\bk\cdot\bx}\vert\bk\vert\tilde{f}(\bk) \ , \\
\tilde{f}(\bk) &= \int d^3 x  e^{-i\bk\cdot\bx}f(\bx) \ .
\end{align}
We note that (\ref{H_F}) is equivalent to the usual oscillator representation of $H_F$.

To facilitate the treatment of random media, it will prove convenient to introduce a continuum model of the atomic degrees of freedom~\cite{Kraisler_2022}. The Hamiltonian of the atoms is given by
\begin{align}
H_A = \hbar \Omega\int d^3x\rho(\bx)\sigma^{\dagger}(\bx)\sigma(\bx) \ ,
 \label{HA}
\end{align}
where $\Omega$ is the atomic resonance frequency, $\rho(\bx)$ is the number density of the atoms, and $\sigma$ 
is a Fermi field that obeys the anticommutation relations
\begin{align}
\label{anticommutation}
\{\sigma(\bx),\sigma^{\dagger}(\bx')\}&=\frac{1}{\rho(\bx)}\delta(\bx-\bx') \ , \\
\{\sigma(\bx),\sigma(\bx')\}&=0 \ .
\end{align}

The Hamiltonian describing the interaction between the field and the atoms is taken to be
 \begin{align}
H_I = \hbar \coup\int d^3x \rho(\bx)\left( \phi^\dagger(\bx)\sigma(\bx)+\sigma^{\dagger}(\bx)\phi(\bx)\right) \ ,
\end{align}
where $\coup$ is the strength of the atom-field coupling. Here we have made the Markovian approximation, in which the coupling constant is independent of the frequency of the photons or positions of the atoms, and have imposed the rotating wave approximation (RWA). 


We suppose that the system is in a two-photon state of the form
\begin{align}
\label{state}
\vert\Psi\rangle = \int d^3 x_1 d^3 x_2 & \left(\psi_2(\bx_1,\bx_2,t)\phi^{\dagger}(\bx_1)\phi^{\dagger}(\bx_2)
+ \psi_1(\bx_1,\bx_2,t)\rho(\bx_1)\sigma^{\dagger}(\bx_1)\phi^{\dagger}(\bx_2)\right.\\
+\nonumber&\left. a(\bx_1,\bx_2,t)\rho(\bx_1)\rho(\bx_2)\sigma^{\dagger}(\bx_1)\sigma^{\dagger}(\bx_2)\right)\vert 0\rangle \ ,
\end{align}
where $\vert 0\rangle$ is the combined vacuum state of the field and the ground state of the atoms. Here the atomic amplitude $a(\bx_1,\bx_2,t)$ is the probability amplitude for exciting two atoms at the points $\bx_1$ and $\bx_2$ at time $t$, the one-photon amplitude $\psi_1(\bx_1,\bx_2,t)$ is the probability amplitude for exciting an atom at $\bx_1$ and creating a photon at $\bx_2$, and the two-photon  amplitude $\psi_2(\bx_1,\bx_2,t)$ is the probability amplitude for creating photons at $\bx_1$ and $\bx_2$.  The functions $\psi_2$ and $a$ are symmetric and antisymmetric, respectively:
\begin{equation}
\psi_2(\bx_2,\bx_1,t) = \psi_2(\bx_1,\bx_2,t) \ , \quad a(\bx_2,\bx_1,t) = -a(\bx_1,\bx_2,t) \ ,
\end{equation}
consistent with the bosonic  and fermionic character of the corresponding fields. 

The state $\vert\Psi\rangle$ is the most general two-photon state within the RWA. In addition, $\vert\Psi\rangle$ is normalized so that $\langle \Psi\vert\Psi\rangle = 1$. It follows from (\ref{commutation}) and (\ref{anticommutation}) that the amplitudes obey the normalization condition
\begin{align}
\int d^3 x_1 d^3 x_2 \left(2\vert\psi_2(\bx_1,\bx_2,t)\vert^2 + \rho(\bx_1)\vert \psi_1(\bx_1,\bx_2,t)\vert^2+2\rho(\bx_1)\rho(\bx_2)\vert a(\bx_1,\bx_2,t)\vert^2\right) = 1
\ .
\end{align}

If the amplitudes $a(\bx_1,\bx_2,t)$, $\psi_1(\bx_1,\bx_2,t)$ and $\psi_2(\bx_1,\bx_2,t)$ are factorizable as functions of $\bx_1$ and $\bx_2$, then there are no quantum correlations and the state $\vert\Psi\rangle$ is not entangled. Otherwise $\vert\Psi\rangle$ is entangled. If $\psi_2$ alone is not factorizable, we say that  $\vert\Psi\rangle$ is an entangled two-photon state.

The dynamics of $\vert\Psi\rangle$ is governed by the Schrodinger equation
\begin{align} 
\label{eq:schr}
i\hbar\partial_t\vert\Psi\rangle = H\vert\Psi\rangle \ .
\end{align}
Projecting onto the states $\phi^{\dagger}(\bx)\vert 0\rangle$ and $\sigma^{\dagger}(\bx)\vert 0\rangle$ and making use of (\ref{commutation}) and (\ref{anticommutation}), we arrive at the following system of equations obeyed by $a$, $\psi_1$ and
$\psi_2$:
\begin{align}
\label{eq:dynamics1}
&\nonumber i\partial_t\psi_2(\bx_1,\bx_2,t) =c(-\Delta_{\bx_1})^{1/2}\psi_2(\bx_1,\bx_2,t)+c(-\Delta_{\bx_2})^{1/2}\psi_2(\bx_1,\bx_2,t) \\
&+\frac{g}{2}(\rho(\bx_1)\psi_1(\bx_1,\bx_2,t)+\rho(\bx_2)\psi_1(\bx_2,\bx_1,t)) \ , \\
 & \nonumber\rho(\bx_1)i\partial_t \psi_1(\bx_1,\bx_2,t) = 2\coup\rho(\bx_1)\psi_2(\bx_1,\bx_2,t)+\rho(\bx_1)\left[c(-\Delta_{\bx_2})^{1/2}+\Omega\right]\psi_1(\bx_1,\bx_2,t)\\
\label{eq:dynamics2}
 &-2\coup\rho(\bx_1)\rho(\bx_2)a(\bx_1,\bx_2,t) \ , \\
  &  \rho(\bx_1)\rho(\bx_2)i\partial_t a(\bx_1,\bx_2,t) =\frac{g}{2}\rho(\bx_1)\rho(\bx_2)(\psi_1(\bx_2,\bx_1,t)-\psi_1(\bx_1,\bx_2,t)) + 2\Omega\rho(\bx_1)\rho(\bx_2)\, a(\bx_1,\bx_2,t) \ .
\label{eq:dynamics3}
\end{align}
The overall factors of $\rho(\bx)$ in the above will be cancelled as necessary. The details of the calculation are presented in Appendix A. 

The above model views the atomic degrees of freedom as fermionic. In Appendix D we consider the bosonic case.

\section{Single-Atom Stimulated Emission}
In this section we consider the problem of stimulated emission by a single atom. We assume that the atom is located at the origin and put $\rho(\bx)=\delta(\bx)$. Thus the system (\ref{eq:dynamics1})--(\ref{eq:dynamics3}) becomes
\begin{align}
\nonumber
&   i\partial_t\psi_2(\bx_1,\bx_2,t) =c(-\Delta_{\bx_1})^{1/2}\psi_2(\bx_1,\bx_2,t)+c(-\Delta_{\bx_2})^{1/2}\psi_2(\bx_1,\bx_2,t) \\
&+\frac{g}{2}(\delta(\bx_1)\psi_1(\bx_1,\bx_2,t)+\delta(\bx_2)\psi_1(\bx_2,\bx_1,t))
\label{eq:emission1} \ ,\\
&\delta(\bx_1)i\partial_t \psi_1(\bx_1,\bx_2,t) = 2\coup\delta(\bx_1)\psi_2(\bx_1,\bx_2,t)+\delta(\bx_1)\left[c(-\Delta_{\bx_2})^{1/2}+\Omega\right]\psi_1(\bx_1,\bx_2,t)
\label{eq:emission2} \ ,
\end{align}
where the term $\delta(\bx_1)\delta(\bx_2)a(\bx_1,\bx_2,t)$ does not contribute due to the antisymmetry of $a$.
We assume that initially there is only one photon present in the system, which means that the initial conditions for the amplitudes $\psi_1$ and $\psi_2$ are given by
\begin{align*}
    \psi_1(0,\bx,0) &= e^{i\bk_0\cdot\bx}\ ,\\
    \psi_2(\bx_1,\bx_2,0) &= 0 \ ,
\end{align*}
where $\bk_0$ is the wavevector of the photon. Note that the amplitude of $\psi_1$ is set to unity for convenience.
To proceed, we take the Laplace transform with respect to $t$ and the Fourier transforms with respect to $\bx_1$ and $\bx_2$ of (\ref{eq:emission1}) and (\ref{eq:emission2}). We thus obtain
\begin{align}
\label{psi1_FL}
iz\psi_2(\bk_1,\bk_2,z)&=\left[c\vert\bk_1\vert+c\vert\bk_2\vert\right]\psi_2(\bk_1,\bk_2)+\frac{g}{2}\left[\psi_1(\bk_1,z)+\psi_1(\bk_2,z)\right]  \ ,\\
    i\left[z\psi_1(\bk,z)-\delta(\bk-\bk_0)\right] &= 2\coup\int d^3 k' \psi_2(\bk',\bk,z) +\left[c\vert\bk\vert+\Omega\right]\psi_1(\bk,z) \ .
\label{psi2_FL}
\end{align}
Here we have defined the Laplace transform by
\begin{align}
a(z)=\int_0^{\infty}dt e^{-zt}a(t) \ ,
\end{align}
where we denote a function and its Laplace transform by the same symbol. Solving (\ref{psi1_FL}) and (\ref{psi2_FL}) leads to an integral equation for $\psi_1(\bk,z)$ of the form
\begin{align} 
\label{laplace_inv}
    \psi_1(\bk,z)=\frac{\delta(\bk-\bk_0)}{z+i(c\vert\bk\vert+\Omega)-i\Sigma(\bk,z)}+\frac{i\coup^2}{z+i(c\vert\bk\vert+\Omega)-i\Sigma(\bk,z)}\int d^3 k' \frac{\psi_1(\bk',z)}{(c\vert\bk'\vert+c\vert\bk\vert)-iz} \ ,
\end{align}
where
\begin{align}
    \Sigma(\bk,z)=g^2\int d^3 k' \frac{1}{(c\vert\bk'\vert+c\vert\bk\vert)-iz-i\epsilon} \ ,
\end{align}
with $\epsilon\to 0^+$.
In order to evaluate the integral in (\ref{laplace_inv}), we make the pole approximation 
where we replace $\Sigma(\bk,z)$ with $\Sigma(\bk,-i(\Omega+c\vert\bk\vert))$. We note that this quantity is  independent of $\bk$ and $z$, and so we will denote it by $\Sigma$. For consistency, we also replace $z$ by $-i(\Omega+c\vert\bk\vert)$ under the integral in (\ref{laplace_inv}). Note that this approximation arises in the Wigner-Weisskopf theory of spontaneous emission. In addition, we split $\Sigma$ into its real and imaginary parts:
\begin{align} 
\label{re}
\re\Sigma &=\delta\omega \ ,\\
\label{im}
\im\Sigma &=\Gamma/2 \ .
\end{align}
We can calculate $\Gamma$ and $\delta\omega$ by making use of the identity
\begin{align}
\frac{1}{c\vert\bk\vert-\Omega-i\epsilon} = P \frac{1}{c\vert\bk\vert-\Omega}+i\pi\delta(c\vert\bk\vert-\Omega) \ .
\end{align}
We find that 
\begin{align}
\label{def_Gamma}
\Gamma &= 2g^2\pi\displaystyle\int\frac{d^3 k}{(2\pi)^3}\,\delta(c\vert\bk\vert-\Omega)\\
&=\frac{g^2\Omega^2}{\pi c^3}  \ ,
\end{align}
and
\begin{equation}
\label{def_lamb}
\delta\omega = \frac{g^2}{2\pi^2} \int_0^{2\pi/\Lambda} \frac{k^2 dk}{ck-\Omega} \ ,
\end{equation}
where we have introduced a high-frequency cutoff to regularize the divergent integral. Using the above results,  (\ref{laplace_inv}) becomes
\begin{align}
\label{psi1_iter}
     \psi_1(\bk,z)=\frac{\delta(\bk-\bk_0)}{z+i(c\vert\bk\vert+\Omega)-i\Sigma}+\frac{ig^2}{z+i(c\vert\bk\vert+\Omega)-i\Sigma}\int d^3 k' \frac{\psi_1(\bk',z)}{c\vert\bk'\vert-\Omega} \ .
\end{align}
Since $\psi_1$ appears on both the left- and right-hand sides of (\ref{psi1_iter}), upon iteration we obtain an infinite series for $\psi_1$, which to order $O(g^2)$ is given by
\begin{align}
    \psi_1(\bk,z) &=\frac{\delta(\bk-\bk_0)}{z+i(c\vert\bk\vert+\Omega)-i\Sigma}+\frac{ig^2}{z+i(\Omega-c\vert\bk'\vert)-i\Sigma}\int d^3 k' \frac{\delta(\bk'-\bk_0)}{(\Omega-c\vert\bk'\vert)(z+i(c\vert\bk'\vert+\Omega)-i\Sigma)}\\
    &=\frac{\delta(\bk-\bk_0)}{z+i(c\vert\bk\vert+\Omega)-i\Sigma}+\frac{ig^2}{z+i(c\vert\bk\vert+\Omega)-i\Sigma}\frac{1}{(c\vert\bk_0\vert-\Omega)(z+i(c\vert\bk_0\vert+\Omega)-i\Sigma)}\ .
\end{align}
Inverting the Laplace transform, we find that $\psi_1(\bk,t)$ is given by
\begin{align}
   \psi_1(\bk,t) &=e^{-i(\Omega-\delta\omega)t-\Gamma t/2}\left[\delta(\bk-\bk_0)e^{-ic\vert\bk\vert t} +\frac{i\coup^2}{((c\vert\bk_0\vert-\Omega)(c\vert\bk_0\vert-c\vert\bk\vert)}\left(e^{-ic\vert\bk\vert t} - e^{-ic\vert\bk_0\vert t}\right)\right]\ .
\end{align}
Note that $\psi_1$ decays exponentially at long times ($\Gamma t \gg 1$) and that $\psi_2$ can be obtained from (\ref{psi1_FL}).
 
\section{Constant Density}

\begin{figure}[t]
\centering
\includegraphics{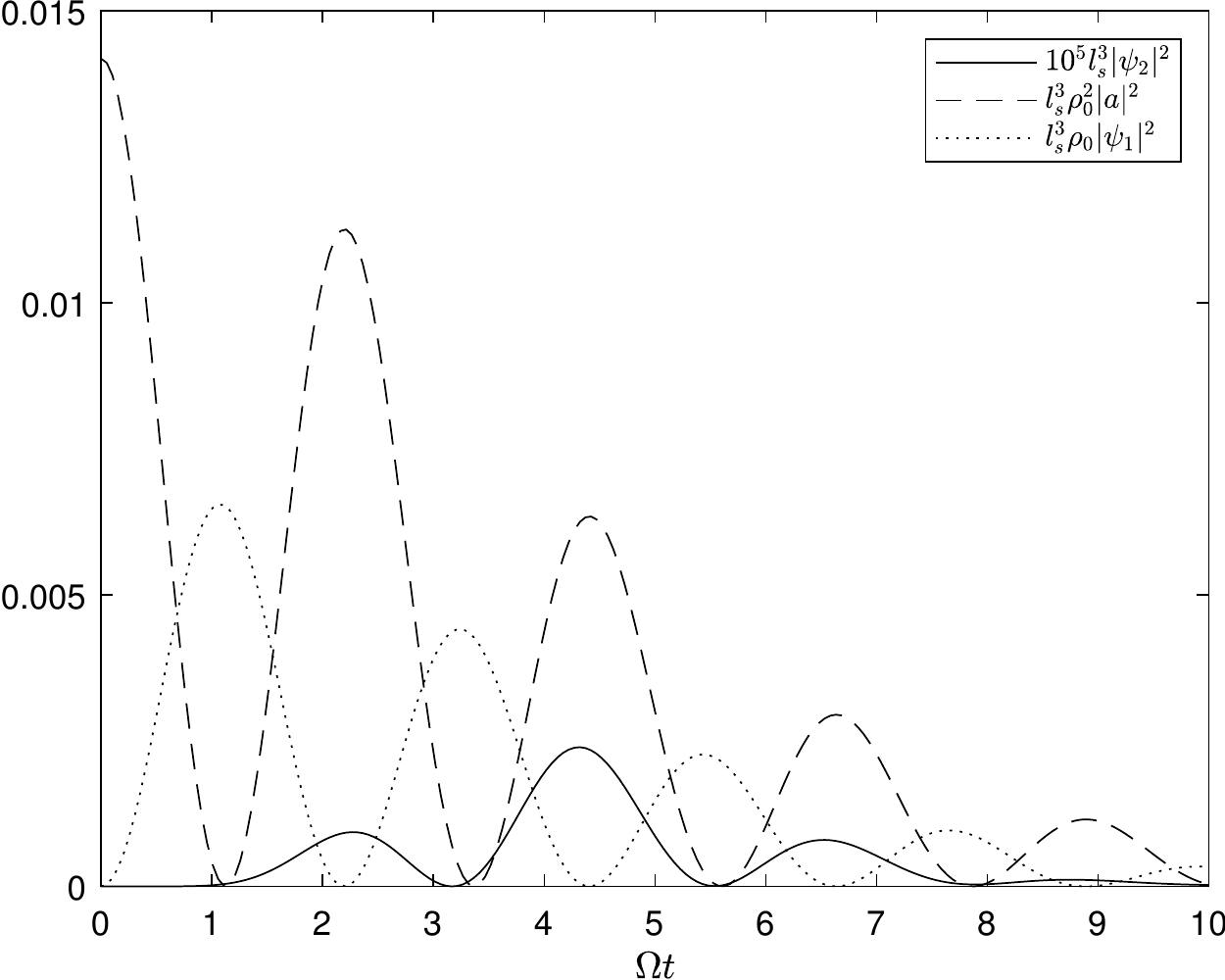}
\caption{Probability densities with distances $\vert\bx_1-\br_1\vert=\vert\bx_2-\br_1\vert=\vert\bx_1-\br_2\vert=\vert\bx_2-\br_2\vert=l_s$}
\label{fig:constant}
\end{figure}

In this section we consider the case of a homogeneous medium and set $\rho(\bx)=\rho_0$, where $\rho_0$ is constant. It is useful to define the function $\tilde{\psi_1}(\bx_1,\bx_2,t)=\psi_1(\bx_2,\bx_1,t)$ and write (\ref{eq:dynamics3}) as a $4\times4$ symmetric system. If we further define the vector 
\begin{align} \label{Psi}
\bPsi(\bx_1,\bx_2,t)=\begin{bmatrix} \sqrt{2}\psi_2(\bx_1,\bx_2,t) \\ \sqrt{\rho_0/2}\psi_1(\bx_1,\bx_2,t) \\ \sqrt{\rho_0/2}\tilde{\psi_1}(\bx_1,\bx_2,t) \\ \sqrt{2}\rho_0 a(\bx_1,\bx_2,t) \end{bmatrix} \ ,
\end{align}
then (\ref{eq:dynamics3}) can be written as
\begin{align} \label{Constant2}
    i\partial_t\bPsi = A\bPsi \ ,
\end{align}
where
\begin{align}
    A=\begin{bmatrix} c(-\Delta_{\bx_1})^{1/2} + c(-\Delta_{\bx_2})^{1/2} & \coup\sqrt{\rho_0} & \coup\sqrt{\rho_0} & 0\\
    \coup\sqrt{\rho_0} & c(-\Delta_{\bx_2})^{1/2}+\Omega & 0 & -\coup\sqrt{\rho_0}\\
    \coup\sqrt{\rho_0} & 0 & c(-\Delta_{\bx_1})^{1/2}+\Omega & \coup\sqrt{\rho_0} \\
    0 & -\coup\sqrt{\rho_0} & \coup\sqrt{\rho_0} & 2\Omega
    \end{bmatrix}.
\end{align}
This definition of $\bPsi$ has the advantage that the matrix $A$ is symmetric. The Fourier transform of (\ref{Constant2}) with respect to the variables $\bx_1$ and $\bx_2$ is given by
\begin{align} \label{Constant3}
    i\partial_t\hat{\bPsi} = \hat{A}\hat{\bPsi}\,
\end{align}
where
\begin{align}\label{Ahatmatrix}
    \hat{A}(\bk_1,\bk_2)=\begin{bmatrix} c\vert\bk_1\vert + c\vert\bk_2\vert & \coup\sqrt{\rho_0} & \coup\sqrt{\rho_0} & 0\\
    \coup\sqrt{\rho_0} & c\vert\bk_2\vert+\Omega & 0 & -\coup\sqrt{\rho_0}\\
    \coup\sqrt{\rho_0} & 0 & c\vert\bk_1\vert+\Omega & \coup\sqrt{\rho_0} \\
    0 & -\coup\sqrt{\rho_0} & \coup\sqrt{\rho_0} & 2\Omega
    \end{bmatrix}.
\end{align}
The matrix $\hat A$ has eigenvalues
\begin{align} \label{evals}
    \nonumber\lambda_1&=b_1 - \frac{\sqrt{b_2-2\sqrt{b_3}}}{2}\, ,\\
    \nonumber\lambda_2&=b_1 + \frac{\sqrt{b_2-2\sqrt{b_3}}}{2}\, ,\\
    \nonumber\lambda_3&=b_1 - \frac{\sqrt{b_2+2\sqrt{b_3}}}{2}\, ,\\
    \nonumber\lambda_4&=b_1 + \frac{\sqrt{b_2+2\sqrt{b_3}}}{2}\, ,
\end{align}
where
\begin{align}
    b_1&=\Omega+\frac{c\,\vert\bk_1\vert}{2}+\frac{c\,\vert\bk_2\vert}{2}\, ,\\
    b_2&=2\,\Omega^2+8\,g^2\rho_0+c^2\,{\vert\bk_1\vert}^2+c^2\,{\vert\bk_2\vert}^2-2\,\Omega\,c\,\vert\bk_1\vert-2\,\Omega\,c\,\vert\bk_2\vert\, ,\\
    \nonumber b_3&=\Omega^4-2\,\Omega^3\,c\,\vert\bk_1\vert-2\,\Omega^3\,c\,\vert\bk_2\vert+\Omega^2\,c^2\,{\vert\bk_1\vert}^2+4\,\Omega^2\,c^2\,\vert\bk_1\vert\,\vert\bk_2\vert+\Omega^2\,c^2\,{\vert\bk_2\vert}^2\\
    \nonumber &+8\,\Omega^2\,g^2\rho_0-2\,\Omega\,c^3\,{\vert\bk_1\vert}^2\,\vert\bk_2\vert-2\,\Omega\,c^3\,\vert\bk_1\vert\,{\vert\bk_2\vert}^2-8\,\Omega\,c\,g^2\rho_0\,\vert\bk_1\vert\\
    &-8\,\Omega\,c\,g^2\rho_0\,\vert\bk_2\vert+c^4\,{\vert\bk_1\vert}^2\,{\vert\bk_2\vert}^2+4\,c^2\,g^2\rho_0\,{\vert\bk_1\vert}^2+4\,c^2\,g^2\rho_0\,{\vert\bk_2\vert}^2.
\end{align}
The components of the associated eigenvectors $\bv_i(\bk_1,\bk_2)$ are given by
\begin{align} \label{evects}
    \nonumber v_{i1} & = -\frac{2\,\Omega^3-2\,\Omega\,g^2\rho_0+2\,\Omega^2\,c\,\vert\bk_1\vert+2\,\Omega^2\,c\,\vert\bk_2\vert-c\,g^2\rho_0\,\vert\bk_1\vert-c\,g^2\rho_0\,\vert\bk_2\vert+2\,\Omega\,c^2\,\vert\bk_1\vert\,\vert\bk_2\vert}{g^2\rho_0\,\left(c\,\vert\bk_1\vert-c\,\vert\bk_2\vert\right)}\\
    \nonumber &-\frac{\left(4\,\Omega+c\,\vert\bk_1\vert+c\,\vert\bk_2\vert\right) \lambda_i^2} {g^2\rho_0\,\left(c\,\vert\bk_1\vert-c\,\vert\bk_2\vert\right)}+\frac{\lambda_i \left(5\,\Omega^2-2\,g^2\rho_0+3\,\Omega\,c\,\vert\bk_1\vert+3\,\Omega\,c\,\vert\bk_2\vert+c^2\,\vert\bk_1\vert\,\vert\bk_2\vert\right)}{g^2\rho_0\,\left(c\,\vert\bk_1\vert-c\,\vert\bk_2\vert\right)}\\
    &+\frac{\lambda_i^3}{g^2\rho_0\,\left(c\,\vert\bk_1\vert-c\,\vert\bk_2\vert\right)}\, ,\\
    v_{i2} & = \frac{\lambda_i^2}{g\sqrt{\rho_0}\,\left(c\,\vert\bk_1\vert-c\,\vert\bk_2\vert\right)}+\frac{2\,\left(\Omega^2+c\,\vert\bk_1\vert\,\Omega-g^2\rho_0\right)}{g\sqrt{\rho_0}\,\left(c\,\vert\bk_1\vert-c\,\vert\bk_2\vert\right)}-\frac{\left(3\,\Omega+c\,\vert\bk_1\vert\right)\,\left(\lambda_i\right)}{g\sqrt{\rho_0}\,\left(c\,\vert\bk_1\vert-c\,\vert\bk_2\vert\right)}\, ,\\
    v_{i3} & = \frac{\lambda_i^2}{g\sqrt{\rho_0}\,\left(c\,\vert\bk_1\vert-c\,\vert\bk_2\vert\right)}+\frac{2\,\left(\Omega^2+c\,\vert\bk_2\vert\,\Omega-g^2\rho_0\right)}{g\sqrt{\rho_0}\,\left(c\,\vert\bk_1\vert-c\,\vert\bk_2\vert\right)}-\frac{\left(3\,\Omega+c\,\vert\bk_2\vert\right)\,\left(\lambda_i\right)}{g\sqrt{\rho_0}\,\left(c\,\vert\bk_1\vert-c\,\vert\bk_2\vert\right)}\, ,\\
    v_{i4} & = 1.
\end{align}
It follows that the solution to (\ref{Constant3}) is
\begin{align}
    \bPsi(\bx_1,\bx_2,t)=\sum_{i=1}^4 \int \frac{d^3 k_1}{(2\pi)^3} \frac{d^3 k_2}{(2\pi)^3} e^{i\bx_1\cdot\bk_1+i\bx_2\cdot\bk_2}C_{i}(\bk_1,\bk_2)e^{-i\coup\sqrt{\rho_0}\lambda_{i}(\bk_1,\bk_2)t}\bv_{i}(\bk_1,\bk_2)\, ,
\end{align}
where the values $C_{i}$ are solutions to the linear system
\begin{align}
    \hat{\bPsi}(\bk_1,\bk_2,0)=\sum_{i=1}^4 \bv_{i}(\bk_1,\bk_2) C_{i}(\bk_1,\bk_2).
\end{align}
In order to study the emission of photons, we assume that initially there are two localized volumes of excited atoms of linear size $l_s$ centered at the points $\br_1$ and $\br_2$. The initial amplitudes are taken to be
\begin{align}
    \psi_2(\bx_1,\bx_2,0) & = 0\, ,\\
    \psi_1(\bx_1,\bx_2,0) & = 0\, ,\\
    a(\bx_1,\bx_2,0) & = \left(\frac{1}{\pi l_s^2}\right)^{3/2}\left(e^{-\vert\bx_1-\br_1\vert^2/2l_s^2}e^{-\vert\bx_2-\br_2\vert^2/2l_s^2}-e^{-\vert\bx_2-\br_1\vert^2/2l_s^2}e^{-\vert\bx_1-\br_2\vert^2/2l_s^2}\right).
\end{align}
Figure~\ref{fig:constant} illustrates the time dependence  of $\psi_1, \psi_2$ and $a$, where we have set the dimensionless quantities ${\Omega}/{\sqrt{\rho_0}\coup}={c}/{l_s\sqrt{\rho_0}\coup}=1$. Note that the atomic amplitude is antisymmetric, consistent with (\ref{anticommutation}) and that it corresponds to an entangled state. We observe that the amplitudes oscillate and decay in time. Moreover, the atomic and one-photon amplitudes are out of phase with one another,  and the two-photon amplitude is an order of magnitude smaller.

\section{Stimulated Emission in Random Media}
\subsection{Kinetic equations}

In this section we consider stimulated emission in a random medium. We suppose that there is at most one atomic excitation and that there are at most two photons present in the field. Thus we set the $a(\bx_1,\bx_2,t)=0$ and 
study the dynamics of $\psi_1$ and $\psi_2$. The system (\ref{eq:dynamics3}) then becomes
\begin{align}
&\nonumber i\partial_t\psi_2(\bx_1,\bx_2,t) =c(-\Delta_{\bx_1})^{1/2}\psi_2(\bx_1,\bx_2,t)+c(-\Delta_{\bx_2})^{1/2}\psi_2(\bx_1,\bx_2,t) \\
&\nonumber+\frac{g}{2}(\rho(\bx_1)\psi_1(\bx_1,\bx_2,t)+\rho(\bx_2)\psi_1(\bx_2,\bx_1,t))\, , \\
 & \nonumber
 i\partial_t \psi_1(\bx_1,\bx_2,t) = 2\coup\psi_2(\bx_1,\bx_2,t)+ \left[c(-\Delta_{\bx_2})^{1/2}+\Omega\right]\psi_1(\bx_1,\bx_2,t)\, ,\\
    &0=\psi_1(\bx_2,\bx_1,t)-\psi_1(\bx_1,\bx_2,t) \ ,
\end{align}
which can be rewritten as the pair of equations
\begin{align}
  \nonumber  i\partial_t\psi_2(\bx_1,\bx_2,t) &=c(-\Delta_{\bx_1})^{1/2}\psi_2(\bx_1,\bx_2,t)+c(-\Delta_{\bx_2})^{1/2}\psi_2(\bx_1,\bx_2,t) \\
  \nonumber &+\frac{g}{2}(\rho(\bx_1)+\rho(\bx_2))\psi_1(\bx_1,\bx_2,t))\, , \\
  i\partial_t \psi_1(\bx_1,\bx_2,t) &= 2\coup\psi_2(\bx_1,\bx_2,t)+\left[\frac{c}{2}(-\Delta_{\bx_1})^{1/2}+\frac{c}{2}(-\Delta_{\bx_2})^{1/2}+\Omega\right]\psi_1(\bx_1,\bx_2,t) \ .
\end{align}
We will assume that the number density $\rho(\bx)$ is of the form
\begin{align} \label{density}
    \rho(\bx)=\rho_0(1+\eta(\bx))\, ,
\end{align}
where $\rho_0$ is a constant and $\eta$ is a real-valued random field. We assume that the correlations of $\eta$ are given by
\begin{align}
    \langle\eta(\bx)\rangle &= 0\, ,\\
    \langle\eta(\bx_1)\eta(\bx_2)\rangle &= C(\bx_1-\bx_2).
\end{align}
where $C$ is the two-point correlation function and $\langle\cdots\rangle$ denotes statistical averaging. We further assume that the medium is statistically homogeneous and isotropic, so that $C$ depends only upon the quantity $\vert\bx-\by\vert$. If we define
\begin{align} 
\bPsi(\bx_1,\bx_2,t)=\begin{bmatrix} \sqrt{2}\psi_2(\bx_1,\bx_2,t)\\ \sqrt{\rho_0}\psi_1(\bx_1,\bx_2,t)\end{bmatrix}
\end{align}
then the above system of equation can be rewritten as
\begin{align} \label{stimulatedsystem}
    i\partial_t\bPsi(\bx_1,\bx_2,t)=A(\bx_1,\bx_2)\bPsi(\bx_1,\bx_2,t) + \coup\sqrt{\frac{\rho_0}{2}}(\eta(\bx_1)+\eta(\bx_2))K\bPsi(\bx_1,\bx_2,t)\, , 
\end{align}
where
\begin{align}
    A(\bx_1,\bx_2)&=\begin{bmatrix}c(-\Delta_{\bx_1})^{1/2}+c(-\Delta_{\bx_2})^{1/2}  & \sqrt{2}\coup\sqrt{\rho_0} \\ \sqrt{2}\coup\sqrt{\rho_0} &\frac{c}{2}(-\Delta_{\bx_1})^{1/2}+\frac{c}{2}(-\Delta_{\bx_2})^{1/2}+\Omega \end{bmatrix}\, , \\
    K &=\begin{bmatrix} 0 & 1 \\ 0 & 0 \end{bmatrix}.
\end{align}
 
To derive a kinetic equation in the high-frequency limit, we rescale the variables $t\to t/\epsilon$, $\bx_1\to\bx_1/\epsilon$, and $\bx_2\to\bx_2/\epsilon$. Additionally, we assume that the randomness is sufficiently weak so that the correlation function $C$ is $O(\epsilon)$.  Eq.~(\ref{stimulatedsystem}) thus becomes
\begin{align}
     i\epsilon\partial_t\bPsi_{\epsilon}(\bx_1,\bx_2,t)=A_{\epsilon}(\bx_1,\bx_2)\bPsi_{\epsilon}(\bx_1,\bx_2,t) + \sqrt{\epsilon}\coup\sqrt{\frac{\rho_0}{2}}(\eta(\bx_1/\epsilon)+\eta(\bx_2/\epsilon))K\bPsi_{\epsilon}(\bx_1,\bx_2,t)\, ,
\end{align}
where
\begin{align}
    A_{\epsilon}(\bx_1,\bx_2)=\begin{bmatrix}\epsilon c(-\Delta_{\bx_1})^{1/2}+\epsilon c(-\Delta_{\bx_2})^{1/2}  & \sqrt{2}\coup\sqrt{\rho_0} \\ \sqrt{2}\coup\sqrt{\rho_0} & \epsilon\frac{c}{2}(-\Delta_{\bx_1})^{1/2}+ \epsilon\frac{c}{2}(-\Delta_{\bx_2})^{1/2}+\Omega \end{bmatrix}.
\end{align}
Next we introduce the scaled Wigner transform, which provides a phase space representation of the correlation functions of the various amplitudes. The Wigner transform $W_{\epsilon}(\bx_1,\bk_1,\bx_2,\bk_2,t)$ is defined by
\begin{align}\label{wignertransform}
    \nonumber &W_{\epsilon}(\bx_1,\bk_1,\bx_2,\bk_2,t)\\
    =&\int\frac{d^3 x_1'}{(2\pi)^3}\frac{d^3 x_2'}{(2\pi)^3}e^{-i\bk_1\cdot\bx_1'-i\bk_2\cdot\bx_2'}\bPsi_{\epsilon}(\bx_1-\epsilon\bx_1'/2,\bx_2-\epsilon\bx_2'/2,t)\bPsi^\dagger(\bx_1+\epsilon\bx_1'/2,\bx_2+\epsilon\bx_2'/2,t)\, ,
\end{align}
where $\dagger$ denotes the hermitian conjugate. The Wigner transform is real-valued and its diagonal elements are related to the probability densities $\vert\psi_{1\epsilon}\vert^2$ and $\vert \psi_{2\epsilon}\vert^2$ by
\begin{align} \label{amplitudes}
    2\vert{\psi_2}_{\epsilon}(\bx_1,\bx_2,t)\vert^2 = \int d^3 k_1 d^3 k_2 (W_{\epsilon}(\bx_1,\bk_1,\bx_2,\bk_2,t))_{11}\, ,\\
    \rho_0\vert {\psi_1}_{\epsilon}(\bx_1,\bx_2,t)\vert^2 = \int d^3 k_1 d^3 k_2 (W_{\epsilon}(\bx_1,\bk_1,\bx_2,\bk_2,t))_{22} .
\end{align}
The above factor of two is due to the symmetry of the function $\psi_2(\bx_1,\bx_2)$. The off diagonal elements are related to correlations between the amplitudes:
\begin{align}
    \sqrt{2\rho_0}{\psi_2}_{\epsilon}(\bx_1,\bx_2,t){\psi_1}_{\epsilon}^{*}(\bx_1,\bx_2,t) =\int d^3 k_1 d^3 k_2 (W_{\epsilon}(\bx_1,\bk_1,\bx_2,\bk_2,t))_{12}.
\end{align}
As shown in Appendix B, the Wigner transform satisfies the Liouville equation
\begin{align} 
\label{Liouville1}
    \nonumber&i\epsilon\partial_t W_{\epsilon}(\bx_1,\bk_1,\bx_2,\bk_2,t) \\
    \nonumber&= \int\frac{\dd^3 k_1'}{(2\pi)^3}\frac{\dd^3 k_2'}{(2\pi)^3}e^{i\bx_1\cdot\bk_1'+i\bx_2\cdot\bk_2'}\hat{A}(\bk_1-\epsilon\bk_1'/2,\bk_2-\epsilon\bk_2'/2)\hat{W}_{\epsilon}(\bk_1',\bk_1,\bk_2',\bk_2,t)\\
    \nonumber&- \int\frac{\dd^3 k_1'}{(2\pi)^3}\frac{\dd^3 k_2'}{(2\pi)^3}e^{i\bx_1\cdot\bk_1'+i\bx_2\cdot\bk_2'}\hat{W}_{\epsilon}(\bk_1',\bk_1,\bk_2',\bk_2,t)\hat{A}(\bk_1+\epsilon\bk_1'/2,\bk_2+\epsilon\bk_2'/2)\\
    \nonumber&+\sqrt{\epsilon}\coup\sqrt{\frac{\rho_0}{2}}\int\frac{\dd^3 q}{(2\pi)^3}e^{i\bq\cdot\bx_1/\epsilon}\,\hat{\eta}(\bq)\left[KW_{\epsilon}(\bx_1,\bk_1+\bq/2,\bx_2,\bk_2,t)-W_{\epsilon}(\bx_1,\bk_1-\bq/2,\bx_2,\bk_2,t)K^{T}\right]\\
    &+\sqrt{\epsilon}\coup\sqrt{\frac{\rho_0}{2}}\int\frac{\dd^3 q}{(2\pi)^3}e^{i\bq\cdot\bx_2/\epsilon}\,\hat{\eta}(\bq)\left[KW_{\epsilon}(\bx_1,\bk_1,\bx_2,\bk_2+\bq/2,t)-W_{\epsilon}(\bx_1,\bk_1,\bx_2,\bk_2-\bq/2,t)K^{T}\right] \ ,
\end{align}
where
\begin{align}
    \hat{A}(\bk_1,\bk_2)=\begin{bmatrix}c\vert\bk_1\vert+ c\vert\bk_2\vert &\sqrt{2\rho_0}\coup\\\sqrt{2\rho_0}g &(c\vert\bk_1\vert+ c\vert\bk_2\vert)/2+\Omega\end{bmatrix} \ ,
\end{align}
and the Fourier transform $\hat{W}$ is defined as
\begin{align}
    \hat{W}_{\epsilon}(\bk_1',\bk_1,\bk_2',\bk_2,t)=\int d^3 x_1 d^3 x_2 e^{-i\bx_1\cdot\bk_1'-i\bx_2\cdot\bk_2'} W_{\epsilon}(\bx_1,\bk_1,\bx_2,\bk_2,t).
\end{align}
We study the behavior of $W_{\epsilon}$ in the high frequency limit $\epsilon\to 0$. To this end, we introduce a multiscale expasion of the Wigner transform of the form
\begin{align} \label{expansion1}
    \nonumber W_{\epsilon}(\bx_1,\bX_1,\bk_1,\bx_2,\bX_2,\bk_2,t)&= W_0(\bx_1,\bk_1,\bx_2,\bk_2,t)+\sqrt{\epsilon}W_1(\bx_1,\bX_1,\bk_1,\bx_2,\bX_2,\bk_2,t)\\
    &+\epsilon W_2(\bx_1,\bX_1,\bk_1,\bx_2,\bX_2,\bk_2,t)+\cdots\, ,
\end{align}
where $\bX_1=\bx_1/\epsilon$ and $\bX_2=\bx_2/\epsilon$ are fast variables and $W_0$ is assumed to be both deterministic and independent of the fast variables. We treat the variables $\bx_1, \bX_1$ and $\bx_2,\bX_2$ as independent and make the replacements
\begin{align} \label{chainrule}
    \nabla_{\bx_i}\to \nabla_{\bx_i}+\frac{1}{\epsilon}\nabla_{\bX_i} \ , \quad i=1,2 \ .
\end{align}  
Hence the Liouville equation becomes
\begin{align} \label{Liouville2}
    \nonumber&i\epsilon\partial_t W_{\epsilon}(\bx_1,\bX_1,\bk_1,\bx_2,\bX_2,\bk_2,t) \\
    \nonumber&= \int\frac{\dd^3 k_1'}{(2\pi)^3}\frac{\dd^3 k_2'}{(2\pi)^3}\frac{\dd^3 K_1}{(2\pi)^3}\frac{\dd^3 K_2}{(2\pi)^3}e^{i\bx_1\cdot\bk_1'+i\bx_2\cdot\bk_2'+i\bX_1\cdot\bK_1+i\bX_2\cdot\bK_2}\\
    \nonumber&\times\hat{A}(\bk_1-\epsilon\bk_1'/2-\bK_1/2,\bk_2-\epsilon\bk_2'/2-\bK_2/2)\hat{W}_{\epsilon}(\bk_1',\bK_1,\bk_1,\bk_2',\bK_2,\bk_2,t)\\
    \nonumber&- \int\frac{\dd^3 k_1'}{(2\pi)^3}\frac{\dd^3 k_2'}{(2\pi)^3}\frac{\dd^3 K_1}{(2\pi)^3}\frac{\dd^3 K_2}{(2\pi)^3}e^{i\bx_1\cdot\bk_1'+i\bx_2\cdot\bk_2'+i\bX_1\cdot\bK_1+i\bX_2\cdot\bK_2}\\
    \nonumber&\times\hat{W}_{\epsilon}(\bk_1',\bK_1,\bk_1,\bk_2',\bK_2,\bk_2,t)\hat{A}(\bk_1-\epsilon\bk_1'/2+\bK_1/2,\bk_2-\epsilon\bk_2'/2+\bK_2/2)\\
    &+\sqrt{\epsilon}\coup\sqrt{\frac{\rho_0}{2}}L_1W_{\epsilon}(\bx_1,\bX_1,\bk_1,\bx_2,\bX_2,\bk_2,t)+\sqrt{\epsilon}\coup\sqrt{\frac{\rho_0}{2}}L_2W_{\epsilon}(\bx_1,\bX_1,\bk_1,\bx_2,\bX_2,\bk_2,t)\, ,
\end{align}
where
\begin{align}
   \nonumber L_1W=\int\frac{\dd^3 q}{(2\pi)^3}e^{i\bq\cdot\bX_1}\,\hat{\eta}(\bq)&\left[KW(\bx_1,\bX_1,\bk_1+\bq/2,\bx_2,\bX_2,\bk_2,t)\right.\\
    &\left.-W_{1}(\bx_1,\bX_1,\bk_1-\bq/2,\bx_2,\bX_2,\bk_2,t)K^{T}\right] \ , \\
   \nonumber L_2W=\int\frac{\dd^3 q}{(2\pi)^3}e^{i\bq\cdot\bX_2}\,\hat{\eta}(\bq)&\left[KW(\bx_1,\bX_1,\bk_1,\bx_2,\bX_2,\bk_2+\bq/2,t)\right.\\
    &\left.-W_{1}(\bx_1,\bX_1,\bk_1,\bx_2,\bX_2,\bk_2-\bq/2,t)K^{T} \right] \ ,
\end{align}
and the Fourier transform $\hat{W}_{\epsilon}(\bk_1',\bK_1,\bk_1,\bk_2',\bK_2,\bk_2,t)$ is defined as
\begin{align}\label{4spatialFT}
    \nonumber &\hat{W}_{\epsilon}(\bk_1',\bK_1,\bk_1,\bk_2',\bK_2,\bk_2,t)\\
    &= \int d^3 x_1 d^3 x_2 d^3 X_1 d^3 X_2 e^{-i\bx_1\cdot\bk_1'-i\bX_1\cdot\bK_1-i\bx_2\cdot\bk_2'-i\bX_2\cdot\bK_2}W_{\epsilon}(\bx_1,\bX_1,\bk_1,\bx_2,\bX_2,\bk_2,t)\ .
\end{align}
Substituting (\ref{expansion1}) into (\ref{Liouville2}) and equating terms of the same order in $\sqrt{\epsilon}$ leads to a hierarchy of equations. At $O(1)$ we have
\begin{align} \label{Order0}
    \hat{A}(\bk_1,\bk_2)W_{0}(\bx_1,\bk_1,\bx_2,\bk_2,t)-W_{0}(\bx_1,\bk_1,\bx_1,\bk_2,t)\hat{A}(\bk_1,\bk_2)=0.
\end{align}
Since $\hat{A}$ is symmetric it can be diagonalized by a unitary transformation. The eigenvalues and eigenvectors of $\hat{A}$ are given by
\begin{align}
    \lambda_{\pm}(\bk_1,\bk_2)&=\frac{3d(\bk_1,\bk_2)+\Omega\pm\sqrt{(d(\bk_1,\bk_2)-\Omega)^2+8g^2\rho_0}}{2}\, ,\\
    \bv_{\pm}(\bk_1,\bk_2)&=\frac{1}{\sqrt{(\lambda_{\pm}(\bk_1,\bk_2)-d(\bk_1,\bk_2)-\Omega)^2+2g^2\rho_0}}\begin{bmatrix} \lambda_{\pm}(\bk_1,\bk_2)-d(\bk_1,\bk_2)-\Omega\\ \coup\sqrt{2\rho_0}\end{bmatrix} \ ,
\end{align}
where
\begin{align}
    d(\bk_1,\bk_2)=\frac{c(\vert\bk_1\vert+\vert\bk_2\vert)}{2} \ .
\end{align}
Evidently $W_0$ is also diagonal in this basis and can be expanded as
\begin{align} \label{changeofbasis1}
    W_0(\bx_1,\bk_1,\bx_2,\bk_2,t)=a_+(\bx_1,\bk_1,\bx_2,\bk_2,t)\bv_+(\bk_1,\bk_2)\bv_+^T(\bk_1,\bk_2)+a_-(\bx_1,\bk_1,\bx_2,\bk_2,t)\bv_-(\bk_1,\bk_2)\bv_-^T(\bk_1,\bk_2).
\end{align}
At order $O(\sqrt{\epsilon})$ we have
\begin{align} \label{Order1}
    \nonumber&\hat{A}(\bk_1-\bK_1/2,\bk_2-\bK_2/2)\hat{W}_{1}(\bx_1,\bK_1,\bk_1,\bx_2,\bK_2,\bk_2)\\
    \nonumber&- \hat{W}_{1}(\bk_1',\bK_1,\bk_1,\bk_2',\bK_2,\bk_2)\hat{A}(\bk_1+\bK_2/2,\bk_2+\bK_2/2)\\
   \nonumber&=\coup\sqrt{\frac{\rho_0}{2}}(2\pi)^3\left(\hat{\eta}(\bK_1)\delta(\bK_2)+\hat{\eta}(\bK_2)\delta(\bK_1)\right)\\
   &\times\left[W_{0}(\bx_1,\bk_1-\bK_1/2,\bx_2,\bk_2-\bK_2/2)K^{T}-KW_{0}(\bx_1,\bk_1+\bK_1/2,\bx_2,\bk_2+\bK_2/2)\right].
\end{align}
We can then decompose $\hat{W_1}$ as
\begin{align}
     \nonumber &\hat{W}_1(\bx_1,\bK_1,\bk_1,\bx_2,\bK_2,\bk_2)\\
      &=\sum_{i,j} w_{i,j}(\bx_1,\bK_1,\bk_1,\bx_2,\bK_2,\bk_2)\bv_{i}(\bk_1-\bK_1/2,\bk_2-\bK_2/2)\bv_{j}^{T}(\bk_1+\bK_1/2,\bk_2+\bK_2/2).
\end{align}
Multiplying (\ref{Order1}) on the left by $\bv_{m}^{T}(\bk_1-\bK_1/2,\bk_2-\bK_2/2)$ and the right by $\bv_{n}(\bk_1+\bK_1/2,\bk_2+\bK_2/2)$, we arrive at
\begin{align} \label{eq:wmn}
    \nonumber&(\lambda_{m}(\bk_1-\bK_1/2,\bk_2-\bK_2/2)-\lambda_{n}(\bk_1+\bK_1/2,\bk_2+\bK_2/2)+i\theta)w_{m,n}(\bx_1,\bK_1,\bk_1,\bx_2,\bK_2,\bk_2)\\
    =\nonumber&\coup\sqrt{\frac{\rho_0}{2}}(2\pi)^3\left\{\eta(\bK_1)\delta(\bK_2)+\eta(\bK_2)\delta(\bK_1)\right\}\\
    \nonumber&\times\left[a_{m}(\bk_1-\bK_1/2,\bk_2-\bK_2/2)K_{m,n}(\bk_1-\bK_1/2,\bk_2-\bK_2/2,\bk_1+\bK_1/2,\bk_2+\bK_2/2)\right]\\
    -\nonumber&\coup\sqrt{\frac{\rho_0}{2}}(2\pi)^3\left\{\eta(\bK_1)\delta(\bK_2)+\eta(\bK_2)\delta(\bK_1)\right\}\\
    &\times\left[a_{n}(\bk_1+\bK_1/2,\bk_2+\bK_2/2)K_{m,n}(\bk_1+\bK_1/2,\bk_2+\bK_2/2,\bk_1-\bK_1/2,\bk_2-\bK_2/2)\right]\, ,
\end{align}
where $\theta\to 0$ is a small positive regularizing parameter and 
\begin{align}
    &\nonumber K_{m,n}(\bk_1,\bk_2,\bq_1,\bq_2)\\
    \nonumber&=\bv_{m}^{T}(\bk_1,\bk_2)K\bv_{n}(\bq_1,\bq_2)\\
    &=\frac{\coup\sqrt{2\rho_0}(\lambda_{m}(\bk_1,\bk_2)-d(\bk_1,\bk_2)-\Omega) }{\sqrt{(\lambda_{m}(\bk_1,\bk_2)-d(\bk_1,\bk_2)-\Omega)^2+2\coup^2\rho_0}\sqrt{(\lambda_{n}(\bq_1,\bq_2)-d(\bq_1,\bq_2)-\Omega)^2+2\coup^2\rho_0}}.
\end{align}
At order $O(\epsilon)$ we find that
\begin{align}\label{Order2}
    \nonumber&i\partial_t W_{0}(\bx_1,\bk_1,\bx_2,\bk_2,t) = LW_2(\bx_1,\bX_1,\bk_1,\bx_2,\bX_2,\bk_2,t)\\
    \nonumber&-M(\bx_1,\bk_1,\bx_2,\bk_2)W_{0}(\bx_1,\bk_1,\bx_2,\bk_2,t)- W_{0}(\bx_1,\bk_1,\bx_2,\bk_2,t)M(\bx_1,\bk_1,\bx_2,\bk_2)\\
    &+L_1W_1+L_2W_1,
\end{align}
where
\begin{align}
    \nonumber &LW_2(\bx_1,\bX_1,\bk_1,\bx_2,\bX_2,\bk_2,t)\\
    \nonumber&=\int\frac{\dd^3 K_1}{(2\pi)^3}\frac{\dd^3 K_2}{(2\pi)^3}e^{i\bX_1\cdot\bK_1+i\bX_2\cdot\bK_2}\left[\hat{A}(\bk_1-\bK_1/2,\bk_2-\bK_2/2)\hat{W}_{2}(\bx_1,\bK_1,\bk_1,\bx_2,\bK_2,\bk_2,t)\right.\\
    \nonumber&\left.-\hat{W}_{2}(\bx_1,\bK_1,\bk_1,\bx_2,\bK_2,\bk_2,t)\hat{A}(\bk_1+\bK_1/2,\bk_2+\bK_2/2)\right]\, ,\\
    &M(\bx_1,\bk_1,\bx_2,\bk_2)=\frac{i}{2}\begin{bmatrix}c\hat\bk_1\cdot\nabla_{\bx_1}+c\hat\bk_2\cdot\nabla_{\bx_2} & 0\\0 &\frac{c}{2}\hat\bk_1\cdot\nabla_{\bx_1}+\frac{c}{2}\hat\bk_2\cdot\nabla_{\bx_2}\end{bmatrix}.
\end{align}
In order to obtain an equation satisfied by $a_\pm$,  we multiply this equation on the left by $\bv_+^T(\bk_1,\bk_2)$ ($\bv_-^T(\bk_1,\bk_2)$) and on the right by $\bv_+(\bk_1,\bk_2)$ ($\bv_-(\bk_1,\bk_2)$) and take the average. Additionally, we assume that $\langle\bv_\pm^T L W_2 \bv_\pm\rangle$ is identically zero, which corresponds to $W_2$ being statistically stationary with respect to the fast variables $\bX_1$ and $\bX_2$. This relation closes the hierarchy of equations and leads to the kinetic equation
\begin{align}\label{transport1}
     \nonumber\frac{1}{c}\partial_t a_{\pm}+f_\pm(\bk_1,\bk_2)\left(\hat\bk_1\cdot\nabla_{\bx_1}+\hat\bk_2\cdot\nabla_{\bx_2}\right)a_{\pm}+\mu_{1\pm}a_{\pm}+\mu_{2\pm}a_{\pm}\\
    =\mu_{1\pm} \int d\hbk'\,A(\hbk_1,\hbk',\vert\bk_1\vert)a_{\pm}(\vert\bk_1\vert\hbk',\bk_2)+\mu_{2\pm}\int d\hbk'\,A(\hbk_2,\hbk',\vert\bk_2\vert)a_{\pm}(\bk_1,\vert\bk_2\vert\hbk')\, ,
\end{align}
where the scattering coefficients $\mu_{i\pm}$, the scattering kernel $A$, and the functions $f_\pm$ are defined as
\begin{align}
    \mu_{1\pm}(\bk_1,\bk_2)&=\frac{g^2\rho_0\pi}{2}\frac{K_{\pm,\pm}(\bk_1,\bk_2,\bk_1,\bk_2)^2}{\frac{c}{4}\left\vert d(\bk_1,\bk_2)-\Omega+\frac{3}{2}\sqrt{(d(\bk_1,\bk_2)-\Omega)^2+8g^2\rho_0}\right\vert}\vert\bk_1\vert^2 \int\frac{d\hbk'}{(2\pi)^3}\,\hat{C}(\vert\bk_1\vert(\hbk_1-\hbk'))\, ,\\
    \mu_{2\pm}(\bk_1,\bk_2)&=\frac{g^2\rho_0\pi}{2}\frac{K_{\pm,\pm}(\bk_1,\bk_2,\bk_1,\bk_2)^2}{\frac{c}{4}\left\vert d(\bk_1,\bk_2)-\Omega+\frac{3}{2}\sqrt{(d(\bk_1,\bk_2)-\Omega)^2+8g^2\rho_0}\right\vert}\vert\bk_2\vert^2
    \int\frac{d\hbk'}{(2\pi)^3}\,\hat{C}(\vert\bk_2\vert(\hbk_2-\hbk'))\, ,\\
    A(\hbk,\hbk',k)&=\frac{\hat{C}(k(\hbk-\hbk'))}{\displaystyle\int d\hbk'\hat{C}(k(\hbk-\hbk'))}\, ,\\
    f_\pm(\bk_1,\bk_2)&=\frac{(\lambda_{\pm}(\bk_1,\bk_2)-d(\bk_1,\bk_2)-\Omega)^2+\coup^2\rho_0}{(\lambda_{\pm}(\bk_1,\bk_2)-d(\bk_1,\bk_2)-\Omega)^2+2\coup^2\rho_0} \ .
\end{align}
Some details of the derivation of ~(\ref{transport1}) are included in Appendix C.

\subsection{Diffusion approximation}
We now consider the diffusion limit of the kinetic equation (\ref{transport1}). The diffusion approximation (DA) to a kinetic equation of the form
\begin{align} \label{transport2}
    \nonumber\frac{1}{c}\partial_t I&+f_1(\vert\bk_1\vert,\vert\bk_2\vert)\hbk_1\cdot\nabla_{\bx_1}I+f_2(|\bk_1|,|\bk_2|)\hbk_2\cdot\nabla_{\bx_2}I+\mu_{1}(\vert\bk_1\vert,\vert\bk_2\vert)I+\mu_{2}(\vert\bk_1\vert,\vert\bk_2\vert)I\\
    \nonumber&=\mu_{1}(\vert\bk_1\vert,\vert\bk_2\vert) \int d\hbk'\,A_1(\hbk_1,\hbk')I(\vert\bk_1\vert\hbk',\bk_2)\\
    &+\mu_{2}(\vert\bk_1\vert,\vert\bk_2\vert)\int d\hbk'\,A_2(\hbk_2,\hbk')I(\bk_1,\vert\bk_2\vert\hbk')\, ,
\end{align}
is obtained by expanding $I$ into spherical harmonics as
\begin{align} \label{sphericalexpansion}
    I = \frac{1}{16\pi^2}u+\frac{3}{16\pi^2}\bJ_1\cdot\hbk_1+\frac{3}{16\pi^2}\bJ_2\cdot\hbk_2
+\cdots\, ,
\end{align}
where
\begin{align}
     u(\bx_1,\bx_2,t)&=\int\dd\hbk_1\dd\hbk_2\,I(\bx_1,\hbk_1,\bx_2,\hbk_2,t)\label{firstangularmoment}\, ,\\
    \bJ_1(\bx_1,\bx_2,t)&=\int\dd\hbk_1\dd\hbk_2\, \hbk_1 I(\bx_1,\hbk_1,\bx_2,\hbk_2,t)\ ,\\
    \bJ_2(\bx_1,\bx_2,t)&=\int\dd\hbk_1\dd\hbk_2\, \hbk_2 I(\bx_1,\hbk_1,\bx_2,\hbk_2,t) \  .
\end{align}
Integrating (\ref{transport2}) with respect to $\hbk_1$ and $\hbk_2$ we arrive at
\begin{align}
\label{eq_u} 
\frac{1}{c}\partial_t u + f_1\nabla_{\bx_1}\cdot\bJ_1 + f_2\nabla_{\bx_2}\cdot\bJ_2=0.
\end{align}
If instead we multiply by $\hbk_1$ and integrate we obtain
\begin{align}
\label{div1}
\frac{1}{c}\partial_t \bJ_1 + f_1\nabla_{\bx_1}\cdot\sigma_1 + f_2\nabla_{\bx_2}\cdot\sigma_3 +\mu_{1}(1-g_1)\bJ_1 = 0\, ,
\end{align}
where
\begin{align}
    g_1&=\int\dd\hbk_1\,\hbk_1\cdot\hbk_1' A_1(\hbk_1,\hbk_1')\ ,
\end{align}
and
\begin{align}
    \sigma_1(\bx_1,\bx_2,t) &=\int\dd\hbk_1\dd\hbk_2\, \hbk_1\otimes\hbk_1 I(\bx_1,\hbk_1,\bx_2,\hbk_2,t) \label{fourthangularmoment}\, ,\\
    \sigma_3(\bx_1,\bx_2,t) &=\int\dd\hbk_1\dd\hbk_2\, \hbk_1\otimes\hbk_2 I(\bx_1,\hbk_1,\bx_2,\hbk_2,t)\label{fifthangularmoment}\ .
\end{align}
Similarly, multiplying (\ref{transport2}) by $\hbk_2$ and carrying out the indicated integrals leads to
\begin{align}
\label{div2}
\frac{1}{c}\partial_t \bJ_2 + f_2  \nabla_{\bx_2}\cdot\sigma_2 + f_2\nabla_{\bx_2}\cdot\sigma_3 +\mu_{2}(1-g_2)\bJ_2 = 0\, ,
\end{align}
where
\begin{align}
 g_2&=\int\dd\hbk_2\,\hbk_2\cdot\hbk_2'A_2(\hbk_2,\hbk_2')\ 
\end{align}
and
\begin{align}
 \sigma_2(\bx_1,\bx_2,t) &=\int\dd\hbk_2\dd\hbk_2\, \hbk_1\otimes\hbk_2 I(\bx_1,\hbk_1,\bx_2,\hbk_2,t)\label{sixthangularmoment}\ .
\end{align}
Next we substitute (\ref{sphericalexpansion}) into (\ref{fourthangularmoment}), (\ref{fifthangularmoment}), and (\ref{sixthangularmoment}) and carry out the indicated integrations. We find that
\begin{align}
    \nabla_{\bx_1}\cdot\sigma_1&=\frac{1}{3}\nabla_{\bx_1} u\ ,\\
    \nabla_{\bx_2}\cdot\sigma_2&=\frac{1}{3}\nabla_{\bx_2} u\ ,\\
    \nabla_{\bx_1}\cdot\sigma_3&=\nabla_{\bx_2}\cdot\sigma_3=0\ .
\end{align}
Using the above results, (\ref{div1}) and (\ref{div2}) become
\begin{align}
\label{J1}
\frac{1}{c}\partial_t \bJ_1  +\mu_{1}(1-g_1)\bJ_1 = - \frac{f_1}{3}\nabla_{\bx_1} u \ , \\
\frac{1}{c}\partial_t \bJ_2  +\mu_{2}(1-g_2)\bJ_2 = - \frac{f_2}{3}\nabla_{\bx_2} u \ .
\label{J2}
\end{align}
At long times ($t\gg 1/[c(1-g_{1,2})\mu_{1,2}]$), the first terms on the right-hand sides of (\ref{J1}) and (\ref{J2}) can be neglected. Substituting the resulting expressions for $\bJ_1$ and $\bJ_2$ into 
(\ref{eq_u}), we obtain the diffusion equation obeyed by $u$:
\begin{align}
\label{diff_eq}
\partial_t u - D_1\Delta_{\bx_1}u - D_2\Delta_{\bx_2}u =0 \ .
\end{align}
Here the diffusion coefficients  are defined by
\begin{align}
D_i=\frac{cf_i^2}{3\mu_{i}(1-g_i)} \ , \quad i=1,2 \ .
\end{align}
In the case of white noise disorder, where the correlation function $C(\bx)=C_0\delta(\bx)$, with constant $C_0$,  the phase functions $A_{1,2}={1}/{4\pi}$, which corresponds to isotropic scattering.

The solution to (\ref{diff_eq}) for an infinite medium is given by
\begin{align}
\label{soln}
u(\bx_1,\bx_2,t) = \frac{1}{(4\pi D_1 t)^{3/2}(4\pi D_2 t)^{3/2}}\int d^3x_1' d^3x_2' \exp\left[-\frac{|\bx_1-\bx_1'|^2}{4D_1 t}  -\frac{|\bx_2-\bx_2'|^2}{4D_2 t} \right]u(\bx_1',\bx_2',0) \ .
\end{align}

Making use of the above results, we see that each of the modes $a_{\pm}$ satisfy (\ref{transport1}) and thus their first angular moments, which are defined by
\begin{align} \label{anglularmoment1}
    u_{\pm}(\bx_1,k_1,\bx_2,k_2,t)=\int d\hbk_1 d\hbk_2 a_{\pm}(\bx_1,k_1\hbk_1,\bx_2,k_2\hbk_2,t)\, ,
\end{align}
satisfy the equations
\begin{align} \label{diffusion1}
     \partial_t u_{\pm} - D_{1,\pm}\Delta_{\bx_1}u_\pm - D_{2,\pm}\Delta_{\bx_2}u_{\pm} =0\, ,
\end{align}
where
\begin{align}
    D_{i,\pm}=\frac{cf_{\pm}^2}{3\mu_{i\pm}} \ .
\end{align}

We assume that initially there are two photons present in the field localized around the points $\br_1$ and $\br_2$ in a volume of linear size $l_s$. The corresponding initial conditions are given by
\begin{align}
     \psi_2(\bx_1,\bx_2,0)&=\frac{1}{ l_s^3}\left[e^{-\vert\bx_1-\br_1\vert^2/2l_s^2}e^{-\vert\bx_2-\br_2\vert^2/2l_s^2}+e^{-\vert\bx_1-\br_2\vert^2/2l_s^2}e^{-\vert\bx_2-\br_1\vert^2/2l_s^2}\right]\  ,\\
    \psi_1(\bx_1,\bx_2,0)&=0 \  .
\end{align}
Note that $\psi_2$ corresponds to an entangled two-photon state.
These initial conditions imply initial conditions for the Wigner transform $W_0$ from (\ref{wignertransform}), which in turn imply initial conditions for the modes $a_{\pm}$ from (\ref{changeofbasis1}):
\begin{align}
    \nonumber &a_{-}(\bx_1,\bk_1,\bx_2,\bk_2,0)\\
    \nonumber &=\frac{1}{\pi^3}{\gamma_{-}(\vert\bk_1\vert,\vert\bk_2\vert)} \left[e^{-\vert\bx_1-\br_1\vert^2/l_s^2}e^{-\vert\bx_2-\br_2\vert^2/l_s^2}+e^{-\vert\bx_1-\br_2\vert^2/l_s^2}e^{-\vert\bx_2-\br_1\vert^2/l_s^2}\right.\\
    \nonumber&+e^{-\vert \bx_1-(\br_1+\br_2)/2\vert^2/l_s^2}e^{-\vert \bx_2-(\br_1+\br_2)/2\vert^2/l_s^2}e^{-i\bk_1\cdot(\br_1-\br_2)-i\bk_2\cdot(\br_2-\br_1)}\\
    &\left.+e^{-\vert \bx_1-(\br_1+\br_2)/2\vert^2/l_s^2}e^{-\vert \bx_2-(\br_1+\br_2)/2\vert^2/l_s^2}e^{-i\bk_1\cdot(\br_2-\br_1)-i\bk_2\cdot(\br_1-\br_2)}\right]\, ,
\end{align}
\begin{align} 
    \nonumber &a_{+}(\bx_1,\bk_1,\bx_2,\bk_2,0)\\
    \nonumber &=\frac{1}{\pi^3}\gamma_{+}(\vert\bk_1\vert,\vert\bk_2\vert)\left[e^{-\vert\bx_1-\br_1\vert^2/l_s^2}e^{-\vert\bx_2-\br_2\vert^2/l_s^2}+e^{-\vert\bx_1-\br_2\vert^2/l_s^2}e^{-\vert\bx_2-\br_1\vert^2/l_s^2}\right.\\
    \nonumber&+e^{-\vert \bx_1-(\br_1+\br_2)/2\vert^2/l_s^2}e^{-\vert \bx_2-(\br_1+\br_2)/2\vert^2/l_s^2}e^{-i\bk_1\cdot(\br_1-\br_2)-i\bk_2\cdot(\br_2-\br_1)}\\
    &\left.+e^{-\vert \bx_1-(\br_1+\br_2)/2\vert^2/l_s^2}e^{-\vert \bx_2-(\br_1+\br_2)/2\vert^2/l_s^2}e^{-i\bk_1\cdot(\br_2-\br_1)-i\bk_2\cdot(\br_1-\br_2)}\right]\, ,
\end{align}
where
\begin{align}
    \gamma_{\pm}(k_1,k_2) = \frac{(\lambda_{\pm}(k_1,k_2)-d(k_1,k_2)-\Omega)^2+2g^2\rho_0}{(\lambda_{\pm}(k_1,k_2)-d(k_1,k_2)-\Omega)^2-(\lambda_{\mp}(k_1,k_2)-d(k_1,k_2)-\Omega)^2} e^{-l_s^2 k_1^2-l_s^2 k_2^2}\ .
\end{align}
The initial conditions for the first angular moments $u_{\pm}$ are then found using ~(\ref{anglularmoment1}):

\begin{align} 
\label{initial1}
\nonumber & u_{-}(\bx_1,k_1,\bx_2,k_2,0)\\
\nonumber &=\frac{16\gamma_{-}(k_1,k_2)}{\pi}\left[e^{-\vert\bx_1-\br_1\vert^2/l_s^2}e^{-\vert\bx_2-\br_2\vert^2/l_s^2}+e^{-\vert\bx_1-\br_2\vert^2/l_s^2}e^{-\vert\bx_2-\br_1\vert^2/l_s^2}\right.\\
    &+\left.2e^{-\vert \bx_1-(\br_1+\br_2)/2\vert^2/l_s^2}e^{-\vert \bx_2-(\br_1+\br_2)/2\vert^2/l_s^2}\frac{\sin(k_1\vert\br_1-\br_2\vert)}{k_1\vert\br_1-\br_2\vert}\frac{\sin(k_2\vert\br_1-\br_2\vert)}{k_2\vert\br_1-\br_2\vert}\right]\ .
\end{align}
\begin{align}
    \nonumber &u_{+}(\bx_1,k_1,\bx_2,k_2,0)\\
    \nonumber &=\frac{16\gamma_{+}(k_1,k_2)}{\pi}\left[e^{-\vert\bx_1-\br_1\vert^2/l_s^2}e^{-\vert\bx_2-\br_2\vert^2/l_s^2}+e^{-\vert\bx_1-\br_2\vert^2/l_s^2}e^{-\vert\bx_2-\br_1\vert^2/l_s^2}\right.\\
    &+\left.2e^{-\vert \bx_1-(\br_1+\br_2)/2\vert^2/l_s^2}e^{-\vert \bx_2-(\br_1+\br_2)/2\vert^2/l_s^2}\frac{\sin(k_1\vert\br_1-\br_2\vert)}{k_1\vert\br_1-\br_2\vert}\frac{\sin(k_2\vert\br_1-\br_2\vert)}{k_2\vert\br_1-\br_2\vert}\right]\ , 
\end{align}
The above initial conditions are then inserted into ~(\ref{soln}) and the Gaussian integrals carried out.
This results in the following formulas for the angular moments $u_{\pm}$:
\begin{align}
    \nonumber &u_{-}(\bx_1,k_1,\bx_2,k_2,t)\\
    \nonumber &=\frac{16\gamma_{-}(k_1,k_2)}{\pi}\left(\frac{l_s^2}{l_s^2+4tD_{1,-}}\right)^{3/2}\left(\frac{l_s^2}{l_s^2+4tD_{2,-}}\right)^{3/2}\\
    &\times\left[e^{-\vert\bx_1-\br_1\vert^2/(l_s^2+4tD_{1,-})}e^{-\vert\bx_2-\br_2\vert^2/(l_s^2+4tD_{2,-})}+e^{-\vert\bx_1-\br_2\vert^2/(l_s^2+4tD_{1,-})}e^{-\vert\bx_2-\br_1\vert^2/(l_s^2+4tD_{2,-})}\right.\\
    &\left.+2e^{-\vert \bx_1-(\br_1+\br_2)/2\vert^2/(l_s^2+4tD_{1,-})}e^{-\vert \bx_2-(\br_1+\br_2)/2\vert^2/(l_s^2+4tD_{2,-})}\frac{\sin(k_1\vert\br_1-\br_2\vert)}{k_1\vert\br_1-\br_2\vert}\frac{\sin(k_2\vert\br_1-\br_2\vert)}{k_2\vert\br_1-\br_2\vert}\right] \ ,
\end{align}
\begin{align}
    \nonumber &u_{+}(\bx_1,k_1,\bx_2,k_2,t)\\
    \nonumber &=\frac{16\gamma_{+}(k_1,k_2)}{\pi}\left(\frac{l_s^2}{l_s^2+4tD_{1,+}}\right)^{3/2}\left(\frac{l_s^2}{l_s^2+4tD_{2,+}}\right)^{3/2}\\
    &\times\left[e^{-\vert\bx_1-\br_1\vert^2/(l_s^2+4tD_{1,+})}e^{-\vert\bx_2-\br_2\vert^2/(l_s^2+4tD_{2,+})}+e^{-\vert\bx_1-\br_2\vert^2/(l_s^2+4tD_{1,+})}e^{-\vert\bx_2-\br_1\vert^2/(l_s^2+4tD_{2,+})}\right.\\
     &\left.+2e^{-\vert \bx_1-(\br_1+\br_2)/2\vert^2/(l_s^2+4tD_{1,+})}e^{-\vert \bx_2-(\br_1+\br_2)/2\vert^2/(l_s^2+4tD_{2,+})}\frac{\sin(k_1\vert\br_1-\br_2\vert)}{k_1\vert\br_1-\br_2\vert}\frac{\sin(k_2\vert\br_1-\br_2\vert)}{k_2\vert\br_1-\br_2\vert}\right].
\end{align}
Finally, combining the above with ~(\ref{amplitudes}), (\ref{changeofbasis1}) and (\ref{anglularmoment1}), we see that the average probability densities are given by
\begin{align}
   &\nonumber \rho_0\langle\vert\psi_1(\bx_1,\bx_2,t)\vert^2\rangle\\
  \nonumber &=\frac{32\coup^2\rho_0}{\pi}\sum_{i=\pm}\int_0^{\infty}\int_0^{\infty}dk_1dk_2 \ \eta_{i}(k_1,k_2)\left(\frac{l_s^2}{l_s^2+4tD_{1,\tilde{i}}}\right)^{3/2}\left(\frac{l_s^2}{l_s^2+4tD_{2,\tilde{i}}}\right)^{3/2}\\
    \nonumber&\times\left[e^{-\vert\bx_1-\br_1\vert^2/(l_s^2+4tD_{1,\tilde{i}})}e^{-\vert\bx_2-\br_2\vert^2/(l_s^2+4tD_{2,\tilde{i}})}+e^{-\vert\bx_1-\br_2\vert^2/(l_s^2+4tD_{1,-})}e^{-\vert\bx_2-\br_1\vert^2/(l_s^2+4tD_{2,-})}\right.\\
 &\left.+2e^{-\vert \bx_1-(\br_1+\br_2)/2\vert^2/(l_s^2+4tD_{1,-})}e^{-\vert \bx_2-(\br_1+\br_2)/2\vert^2/(l_s^2+4tD_{2,-})}\frac{\sin(k_1\vert\br_1-\br_2\vert)}{k_1\vert\br_1-\br_2\vert}\frac{\sin(k_2\vert\br_1-\br_2\vert)}{k_2\vert\br_1-\br_2\vert}\right] \label{mixedamp1}\, ,
\end{align}
\begin{align}
  \nonumber& \langle\vert\psi_2(\bx_1,\bx_2,t)\vert^2\rangle\\
  \nonumber&=\frac{16}{\pi}\sum_{i=\pm}\int_0^{\infty}\int_0^{\infty}dk_1dk_2 \ \zeta_{i}(k_1,k_2)\left(\frac{l_s^2}{l_s^2+4tD_{1,-}}\right)^{3/2}\left(\frac{l_s^2}{l_s^2+4tD_{2,-}}\right)^{3/2}\\
   \nonumber&\times\left[e^{-\vert\bx_1-\br_1\vert^2/(l_s^2+4tD_{1,-})}e^{-\vert\bx_2-\br_2\vert^2/(l_s^2+4tD_{2,-})}+e^{-\vert\bx_1-\br_2\vert^2/(l_s^2+4tD_{1,-})}e^{-\vert\bx_2-\br_1\vert^2/(l_s^2+4tD_{2,-})}\right.\\
    \nonumber &\left.+2e^{-\vert \bx_1-(\br_1+\br_2)/2\vert^2/(l_s^2+4tD_{1,-})}e^{-\vert \bx_2-(\br_1+\br_2)/2\vert^2/(l_s^2+4tD_{2,-})}\frac{\sin(k_1\vert\br_1-\br_2\vert)}{k_1\vert\br_1-\br_2\vert}\frac{\sin(k_2\vert\br_1-\br_2\vert)}{k_2\vert\br_1-\br_2\vert}\right] \label{fieldamp1}\ ,
\end{align}
where we have introduced the notation $\tilde{i} = -i$  for $i=\pm$, and the coefficients $\eta_{\pm}$ and $\zeta_{\pm}$ are defined by
\begin{align}
\eta_{\pm}(k_1,k_2) & = \frac{k_1^2k_2^2e^{-l_s^2k_1^2-l_s^2k_2^2}}{(\lambda_{\pm}(k_1,k_2)-d(k_1,k_2)-\Omega)^2-(\lambda_{\mp}(k_1,k_2)-d(k_1,k_2)-\Omega)^2}\ ,\\ 
    \zeta_{\pm}(k_1,k_2) & =\frac{k_1^2 k_2^2e^{-l_s^2k_1^2-l_s^2k_2^2}(\lambda_{+}(k_1,k_2)-d(k_1,k_2)-\Omega)^2}{(\lambda_{+}(k_1,k_2)-d(k_1,k_2)-\Omega)^2-(\lambda_{-}(k_1,k_2)-d(k_1,k_2)-\Omega)^2} \ .
\end{align} 
It is readily seen that long times,  $\langle\vert\psi_1\vert^2\rangle$ and $\langle\vert\psi_2\vert^2\rangle$ decay algebraically according to the formulas
\begin{align}
\label{long_time_1}
 \rho_0\langle\vert\psi_1\vert^2\rangle &= \frac{C_3}{t^3}-\frac{C_4(\bx_1,\bx_2,\br_1,\br_2)}{t^{4}} \ , \\
 \langle\vert\psi_2\vert^2\rangle &= \frac{C_1}{t^3}-\frac{C_2(\bx_1,\bx_2,\br_1,\br_2)}{t^{4}} \ ,
\label{long_time_2}
\end{align}
where the constants $C_i$ for $i=1,2,3,4$ are given by
\begin{align}
   \nonumber C_1 =
   \sum_{i=\pm}\frac{64}{\pi}\int_0^{\infty}\int_0^{\infty}& dk_1dk_2 \ \zeta_{i}(k_1,k_2)\left(\frac{l_s^2}{4D_{1,\tilde{i}}}\right)^{3/2}\left(\frac{l_s^2}{4D_{2,\tilde{i}}}\right)^{3/2} \\
   &\times\left[1+\frac{\sin(k_1\vert\br_1-\br_2\vert)}{k_1\vert\br_1-\br_2\vert}\frac{\sin(k_2\vert\br_1-\br_2\vert)}{k_2\vert\br_1-\br_2\vert}\right]\ ,
\end{align}
\begin{align}
    \nonumber C_ 2 & =\sum_{i=\pm}\frac{16}{\pi }\int_0^{\infty}\int_0^{\infty}dk_1dk_2\ \zeta_i(k_1,k_2) \left(\frac{l_s^2}{4D_{1,\tilde{i}}}\right)^{3/2}\left(\frac{l_s^2}{4D_{2,\tilde{i}}}\right)^{3/2} \\
    &\times\left[\frac{\vert\bx_1-\br_1\vert^2}{4D_{1,i}}+\frac{\vert\bx_2-\br_2\vert^2}{4D_{2,i}}+\frac{\vert\bx_1-\br_2\vert^2}{4D_{1,i}}+\frac{\vert\bx_2-\br_1\vert^2}{4D_{2,i}}\right. \\
    \nonumber &\left. +2\left(\frac{\vert\bx_1-(\br_1+\br_2)/2\vert^2}{4D_{1,i}}+\frac{\vert\bx_2-(\br_1+\br_2)/2\vert^2}{4D_{2,i}}\right)\frac{\sin(k_1\vert\br_1-\br_2\vert)}{k_1\vert\br_1-\br_2\vert}\frac{\sin(k_2\vert\br_1-\br_2\vert)}{k_2\vert\br_1-\br_2\vert}\right]\ ,
\end{align}
\begin{align}
   \nonumber C_3 =\sum_{i=\pm}\frac{32\coup^2\rho_0}{\pi}\int_0^{\infty}\int_0^{\infty}& dk_1dk_2\ \eta_{i}(k_1,k_2)\left(\frac{l_s^2}{4D_{1,\tilde{i}}}\right)^{3/2}\left(\frac{l_s^2}{4D_{2,\tilde{i}}}\right)^{3/2}\\
   &\times\left[2+2\frac{\sin(k_1\vert\br_1-\br_2\vert)}{k_1\vert\br_1-\br_2\vert}\frac{\sin(k_2\vert\br_1-\br_2\vert)}{k_2\vert\br_1-\br_2\vert}\right]\ ,
\end{align}
\begin{align}
   \nonumber C_4  &=\sum_{i=\pm}\frac{32\coup^2\rho_0}{\pi}\int_0^{\infty}\int_0^{\infty}dk_1dk_2 k_1^2\ \eta_{i}(k_1,k_2) \left(\frac{l_s^2}{4D_{1,i}}\right)^{3/2}\left(\frac{l_s^2}{4D_{2,i}}\right)^{3/2} \\
    \nonumber &\times\left[\frac{\vert\bx_1-\br_1\vert^2}{4D_{1,i}}+\frac{\vert\bx_2-\br_2\vert^2}{4D_{2,i}}+\frac{\vert\bx_1-\br_2\vert^2}{4D_{1,i}}+\frac{\vert\bx_2-\br_1\vert^2}{4D_{2,i}}\right. \\
    & \left. +2\left(\frac{\vert\bx_1-(\br_1+\br_2)/2\vert^2}{4D_{1,i}}+\frac{\vert\bx_2-(\br_1+\br_2)/2\vert^2}{4D_{2,i}}\right)\frac{\sin(k_1\vert\br_1-\br_2\vert)}{k_1\vert\br_1-\br_2\vert}\frac{\sin(k_2\vert\br_1-\br_2\vert)}{k_2\vert\br_1-\br_2\vert}\right] \ .
\end{align}

\begin{figure}[t]
    \centering
    \includegraphics{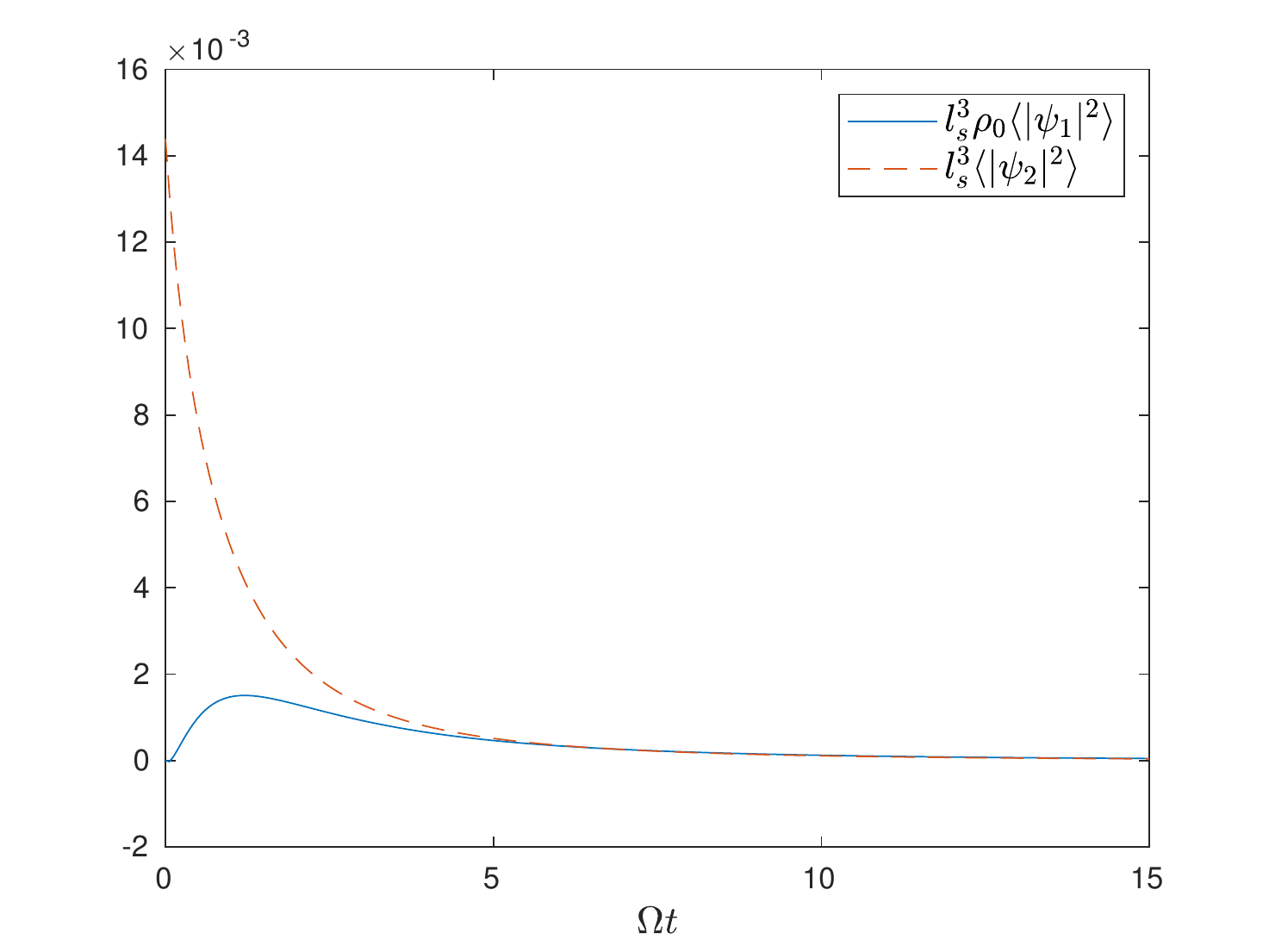}
    \caption{One- and two-photon probability densities for stimulated emission in a random medium. }
    \label{fig:a0amplitudes}
\end{figure}

\begin{figure}[t]
    \centering
    \includegraphics{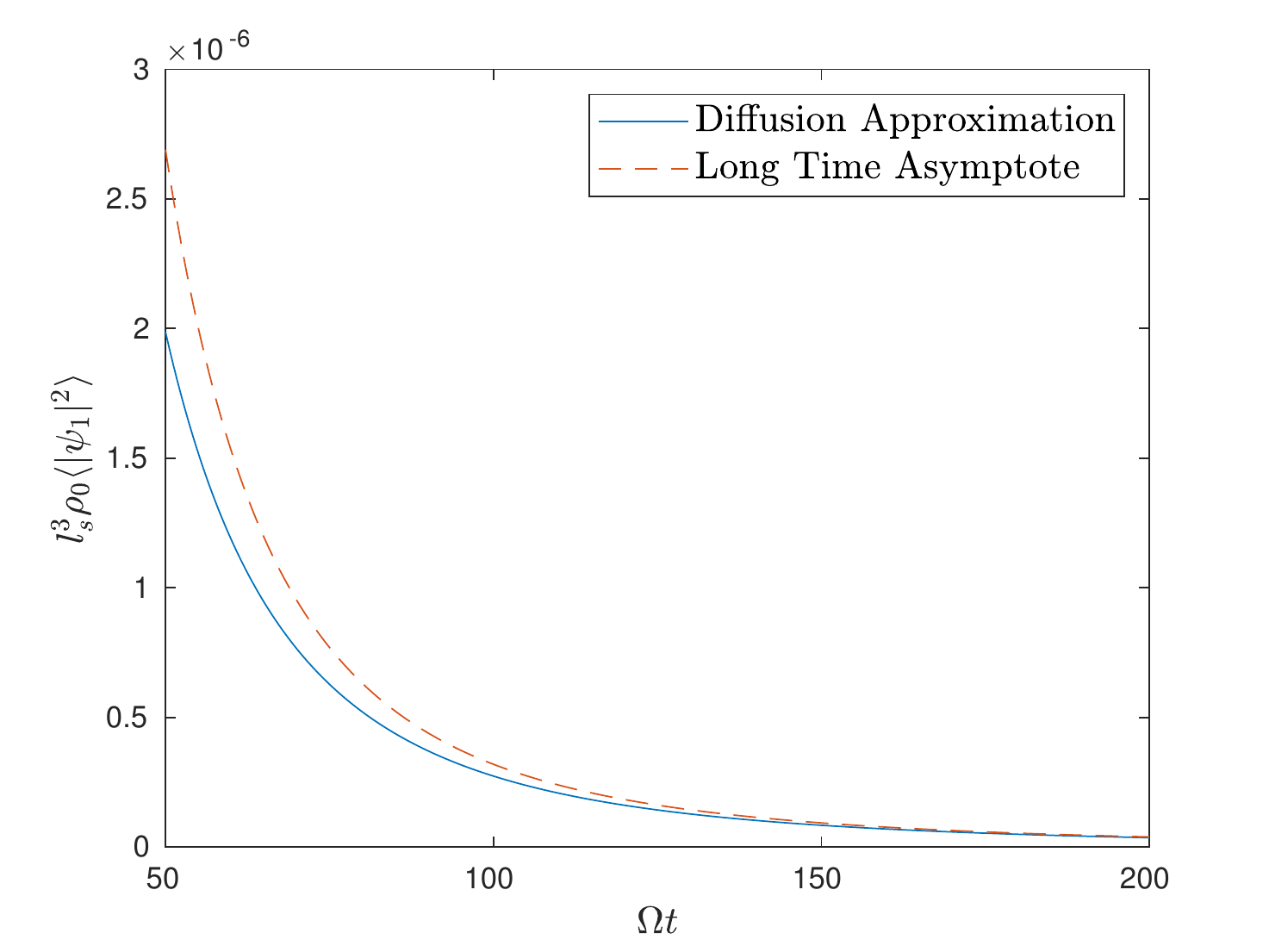}
    \caption{Comparison of diffusion approximation and long-time asymptote for $\langle\vert\psi_1\vert^2\rangle$.}
    \label{fig:a0mixedasymptote}
\end{figure}

\begin{figure}[t]
    \centering
    \includegraphics{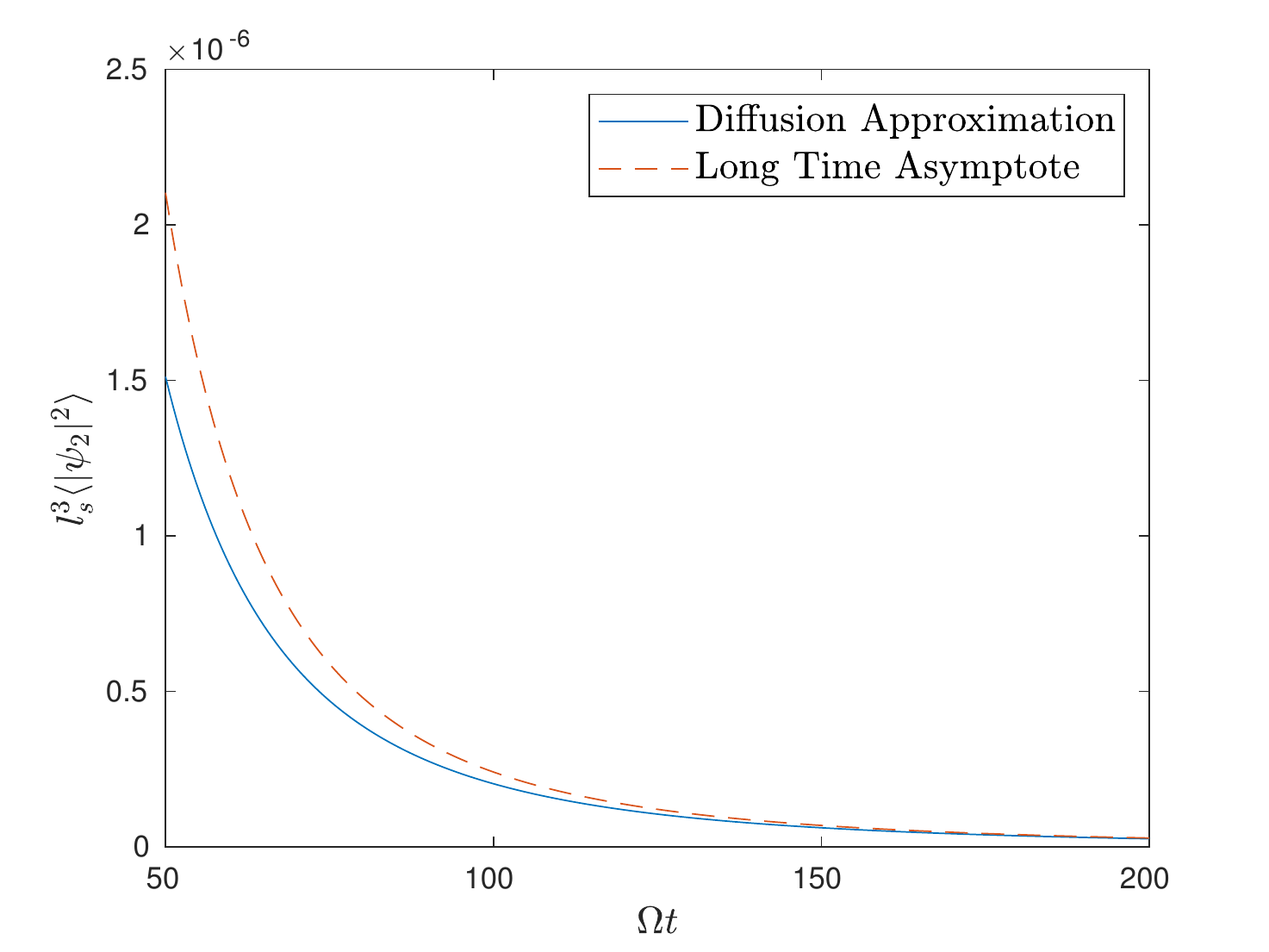}
    \caption{Comparison of diffusion approximation and long-time asymptote for $\langle\vert\psi_2\vert^2\rangle$.}
    \label{fig:a0fieldasymptote}
\end{figure}

To illustrate the above results, we consider the case of isotropic scattering and set the dimensionless quantities ${\Omega}/{\sqrt{\rho_0}\coup}={c}/{l_s\sqrt{\rho_0}\coup}=1$. In addition, we choose  $\bx_1=(l_s,0,0)$, $\bx_2=(-l_s,0,0)$, $\br_1=(0,0,l_s)$ and $\br_2=(0,0,-l_s)$, so that the distances from the points of excitation ($\br_1$ and $\br_2$) to the points of detection are  equal to $l_s$. In Figure~\ref{fig:a0amplitudes} we plot the time dependence of  the probability densities $|\langle \psi_1 \rangle |^2$ and $|\langle \psi_2 \rangle |^2$. We note that  $\vert\psi_2\vert^2$ is monotonically decreasing while $\vert\psi_1\vert^2$ has a peak near $\Omega t=1$. A comparison of these results with the asymptotic formulas (\ref{long_time_1}) and  (\ref{long_time_2}) is shown in Figures \ref{fig:a0mixedasymptote} and \ref{fig:a0fieldasymptote} . There is good agreement at long times.

\section{Kinetic Equations for Two-Photon Light}
In this section, we consider the general problem of two photons interacting with a random medium. That is, we will study the the time evolution of the amplitudes $a(\bx_1,\bx_2,t)$, $\psi_1(\bx_1,\bx_2,t)$ and $\psi_2(\bx_1,\bx_2,t)$ and derive the related kinetic equations. We begin with the system (\ref{eq:dynamics3}), where we have canceled overall factors of $\rho$:
\begin{align} 
\label{dynamics}
&\nonumber i\partial_t\psi_2(\bx_1,\bx_2,t) =c(-\Delta_{\bx_1})^{1/2}\psi_2(\bx_1,\bx_2,t)+c(-\Delta_{\bx_2})^{1/2}\psi_2(\bx_1,\bx_2,t)  \\
&\nonumber
+ \frac{g}{2}(\rho(\bx_1)\psi_1(\bx_1,\bx_2,t)+\rho(\bx_2)\psi_1(\bx_2,\bx_1,t)) \ , \\
&\nonumber
i\partial_t \psi_1(\bx_1,\bx_2,t) = 2\coup\psi_2(\bx_1,\bx_2,t)+\left[c(-\Delta_{\bx_2})^{1/2}+\Omega\right]\psi_1(\bx_1,\bx_2,t)\\
&
-2\coup\rho(\bx_2)a(\bx_1,\bx_2,t) \ , \\
&i\partial_t a(\bx_1,\bx_2,t) =\frac{g}{2}(\psi_1(\bx_2,\bx_1,t)-\psi_1(\bx_1,\bx_2,t)) + 2\Omega a(\bx_1,\bx_2,t) \ .
\nonumber
\end{align}
Here the number density $\rho$ is a random field of the form (\ref{density}). Eq.~(\ref{dynamics}) can now be rewritten as 
\begin{align}\label{eq:system}
i\partial_t\bPsi= A(\bx_1,\bx_2)\bPsi+\coup\sqrt{\rho_0}\eta(\bx_1)K_1\bPsi+\coup\sqrt{\rho_0}\eta(\bx_2)K_2\bPsi\, ,
\end{align}
where $\bPsi$ is defined by (\ref{Psi}) and 
\begin{align}
    A(\bx_1,\bx_2)&=\begin{bmatrix} c(-\Delta_{\bx_1})^{1/2} + c(-\Delta_{\bx_2})^{1/2} & \coup\sqrt{\rho_0} & \coup\sqrt{\rho_0} & 0\\
    \coup\sqrt{\rho_0} & c(-\Delta_{\bx_2})^{1/2}+\Omega & 0 & -\coup\sqrt{\rho_0}\\
    \coup\sqrt{\rho_0} & 0 & c(-\Delta_{\bx_1})^{1/2}+\Omega & \coup\sqrt{\rho_0} \\
    0 & -\coup\sqrt{\rho_0} & \coup\sqrt{\rho_0} & 2\Omega
    \end{bmatrix}\, ,\\
    K_1&=\begin{bmatrix} 0 & 1 & 0 & 0\\  0 & 0 & 0 & 0 \\ 0 & 0 & 0 & 1\\ 0 & 0 & 0 & 0 \end{bmatrix}\, , \quad
    K_2=\begin{bmatrix} 0 & 0 & 1 & 0\\  0 & 0 & 0 & -1 \\ 0 & 0 & 0 & 0\\ 0 & 0 & 0 & 0 \end{bmatrix}.
\end{align}
As before. to derive a kinetic equation in the high-frequency limit, we rescale $t,\bx_1,\bx_2$ according to  $t\to t/\epsilon$, $\bx_1\to\bx_1/\epsilon$ and $\bx_2\to\bx_2/\epsilon$. Additionally, we assume that the randomness is sufficiently weak so that the correlations of $\eta$ are $O(\epsilon)$. 
Eq.~(\ref{eq:system}) thus  becomes
\begin{align}
i\epsilon\partial_t\bPsi_{\epsilon}= A_{\epsilon}(\bx_1,\bx_2)\bPsi_{\epsilon}+\sqrt{\epsilon}\coup\sqrt{\rho_0}\eta(\bx_1/\epsilon)K_1\bPsi_{\epsilon}+\sqrt{\epsilon}\coup\sqrt{\rho_0}\eta(\bx_2/\epsilon)K_2\bPsi_{\epsilon}\, ,
\end{align}
where
\begin{align}
    A_{\epsilon}(\bx_1,\bx_2)=\begin{bmatrix} c\epsilon(-\Delta_{\bx_1})^{1/2} + c\epsilon(-\Delta_{\bx_2})^{1/2} & \coup\sqrt{\rho_0} & \coup\sqrt{\rho_0} & 0\\
    \coup\sqrt{\rho_0} & c\epsilon(-\Delta_{\bx_2})^{1/2}+\Omega & 0 & -\coup\sqrt{\rho_0}\\
    \coup\sqrt{\rho_0} & 0 & c\epsilon(-\Delta_{\bx_1})^{1/2}+\Omega & \coup\sqrt{\rho_0} \\
    0 & -\coup\sqrt{\rho_0} & \coup\sqrt{\rho_0} & 2\Omega
    \end{bmatrix}.
\end{align}
To proceed further, we introduce the  $4\times4$ matrix Wigner transform $W_{\epsilon}$ which is  defined by~(\ref{wignertransform}). The diagonal elements of  $W_{\epsilon}$ are related to the probability densities by
\begin{align}
\label{diag_1}
 2\vert\psi_{2\epsilon}(\bx_1,\bx_2,t)\vert^2 = \int d^3 k_1 d^3 k_2 (W_{\epsilon}(\bx_1,\bk_1,\bx_2,\bk_2,t))_{11}\ ,\\
\label{diag_2}
 \frac{\rho_0}{2}\vert \psi_{1\epsilon}(\bx_1,\bx_2,t)\vert^2 = \int d^3 k_1 d^3 k_2 (W_{\epsilon}(\bx_1,\bk_1,\bx_2,\bk_2,t))_{22}\, ,\\
\rho_0^2\vert a_{\epsilon}(\bx_1,\bx_2,t)\vert^2 = \int d^3 k_1 d^3 k_2 (W_{\epsilon}(\bx_1,\bk_1,\bx_2,\bk_2,t))_{44}\, ,
\label{diag_3}
\end{align}
while the off diagonal elements of $W_{\epsilon}$ are related to correlations between the amplitudes. It can be seen by direct calculation that the Wigner transform satisfies the Liouville equation
\begin{align}\label{Liouville3}
    \nonumber&i\epsilon\partial_t W_{\epsilon}(\bx_1,\bk_1,\bx_2,\bk_2,t) \\
    \nonumber &= \int\frac{\dd^3 k_1'}{(2\pi)^3}\frac{\dd^3 k_2'}{(2\pi)^3}e^{i\bx_1\cdot\bk_1'+i\bx_2\cdot\bk_2'}\hat{A}(\bk_1-\epsilon\bk_1'/2,\bk_2-\epsilon\bk_2'/2)\hat{W}_{\epsilon}(\bk_1',\bk_1,\bk_2',\bk_2,t)\\
    \nonumber &- \int\frac{\dd^3 k_1'}{(2\pi)^3}\frac{\dd^3 k_2'}{(2\pi)^3}e^{i\bx_1\cdot\bk_1'+i\bx_2\cdot\bk_2'}\hat{W}_{\epsilon}(\bk_1',\bk_1,\bk_2',\bk_2,t)\hat{A}(\bk_1+\epsilon\bk_1'/2,\bk_2+\epsilon\bk_2'/2)\\
    \nonumber &+\sqrt{\epsilon}\coup\sqrt{\rho_0}\int\frac{\dd^3 q}{(2\pi)^3}e^{i\bq\cdot\bx_1/\epsilon}\,\hat{\eta}(\bq)\left[K_1W_{\epsilon}(\bx_1,\bk_1+\bq/2,\bx_2,\bk_2,t)-W_{\epsilon}(\bx_1,\bk_1-\bq/2,\bx_2,\bk_2,t)K_1^{T}\right]\\
    &+\sqrt{\epsilon}\coup\sqrt{\rho_0}\int\frac{\dd^3 q}{(2\pi)^3}e^{i\bq\cdot\bx_2/\epsilon}\,\hat{\eta}(\bq)\left[K_2W_{\epsilon}(\bx_1,\bk_1,\bx_2,\bk_2+\bq/2,t)-W_{\epsilon}(\bx_1,\bk_1,\bx_2,\bk_2-\bq/2,t)K_2^{T}\right]\ ,
\end{align}
where $\hat{A}$ is given by~(\ref{Ahatmatrix}), and the Fourier transform $\hat{W_\epsilon}$ is defined by
\begin{align}
    \hat{W}_{\epsilon}(\bk_1',\bk_1,\bk_2',\bk_2,t)=\int d^3 x_1 d^3 x_2 e^{-i\bx_1\cdot\bk_1'-i\bx_2\cdot\bk_2'} W_{\epsilon}(\bx_1,\bk_1,\bx_2,\bk_2,t).
\end{align}
Once again we study the behavior of $W_{\epsilon}$ in the high-frequency limit $\epsilon\to 0$ and introduce a multiscale expansion of the Wigner transform of the form
\begin{align} 
\label{expansion}
    \nonumber W_{\epsilon}(\bx_1,\bX_1,\bk_1,\bx_2,\bX_2,\bk_2,t) &= W_0(\bx_1,\bk_1,\bx_2,\bk_2,t)+\sqrt{\epsilon}W_1(\bx_1,\bX_1,\bk_1,\bx_2,\bX_2,\bk_2,t)\\
    &+\epsilon W_2(\bx_1,\bX_1,\bk_1,\bx_2,\bX_2,\bk_2,t)+\cdots\, ,
\end{align}
where $\bX_1=\bx_1/\epsilon$ and $\bX_2=\bx_2/\epsilon$ are fast variables and $W_0$ is both deterministic and independent of the fast variables. We will treat the slow and fast variables $\bx_1$ and $\bX_1$ (respectively $\bx_2$ and $\bX_2$) as independent and make the replacement (\ref{chainrule}). Hence (\ref{Liouville3}) becomes
\begin{align} \label{Liouville4}
    \nonumber &i\epsilon\partial_t W_{\epsilon}(\bx_1,\bX_1,\bk_1,\bx_2,\bX_2,\bk_2,t) \\
    \nonumber &= \int\frac{\dd^3 k_1'}{(2\pi)^3}\frac{\dd^3 k_2'}{(2\pi)^3}\frac{\dd^3 K_1}{(2\pi)^3}\frac{\dd^3 K_2}{(2\pi)^3}e^{i\bx_1\cdot\bk_1'+i\bx_2\cdot\bk_2'+i\bX_1\cdot\bK_1+i\bX_2\cdot\bK_2}\\
    \nonumber&\times\left[\hat{A}(\bk_1-\bK_1/2-\epsilon\bk_1'/2,\bk_2-\bK_2/2-\epsilon\bk_2'/2)\hat{W}_{\epsilon}(\bk_1',\bK_1,\bk_1,\bk_2',\bK_2,\bk_2,t)\right.\\
    \nonumber&-\left.\hat{W}_{\epsilon}(\bk_1',\bK_1,\bk_1,\bk_2',\bK_2,\bk_2,t)\hat{A}(\bk_1+\bK_2/2+\epsilon\bk_1'/2,\bk_2+\bK_2/2+\epsilon\bk_2'/2)\right]\\
    \nonumber &+\sqrt{\epsilon}\coup\sqrt{\rho_0}\int\frac{\dd^3 q}{(2\pi)^3}e^{i\bq\cdot\bx_1/\epsilon}\,\hat{\eta}(\bq)\left[K_1W_{\epsilon}(\bx_1,\bX_1,\bk_1+\bq/2,\bx_2,\bX_2,\bk_2,t)\right.\\
    \nonumber &-W_{\epsilon}(\bx_1,\bX_1,\bk_1-\bq/2,\bx_2,\bX_2,\bk_2,t)K_1^{T}]\\
    \nonumber&+\sqrt{\epsilon}\coup\sqrt{\rho_0}\int\frac{\dd^3 q}{(2\pi)^3}e^{i\bq\cdot\bx_2/\epsilon}\,\hat{\eta}(\bq)\left[K_2W_{\epsilon}(\bx_1,\bX_1,\bk_1,\bx_2,\bX_2,\bk_2+\bq/2,t)\right.\\
     &-W_{\epsilon}(\bx_1,\bX_1,\bk_1,\bx_2,\bX_2,\bk_2-\bq/2,t)K_2^{T}] \ ,
\end{align}
where the Fourier transform $\hat{W}_{\epsilon}(\bk_1',\bK_1,\bk_1,\bk_2',\bK_2,\bk_2)$ is defined by (\ref{4spatialFT}).
Next we substitute~(\ref{expansion}) into~(\ref{Liouville2}) and collect terms at each order of $\sqrt{\epsilon}$. At order $O(1)$ we have
\begin{align} \label{Order0b}
    \hat{A}(\bk_1,\bk_2)W_{0}(\bx_1,\bk_1,\bx_2,\bk_2)-W_{0}(\bx_1,\bk_1,\bx_1,\bk_2)\hat{A}(\bk_1,\bk_2)=0.
\end{align}
Since $\hat{A}$ is symmetric it can be diagonalized. We then define $\{\bv_i(\bk_1,\bk_2),\lambda_i(\bk_1,\bk_2)\}$, $i=1,2,3,4$ be the eigenvector-eigenvalue pairs given by (\ref{evals}) and (\ref{evects}). It follows from (\ref{Order0b}) that $W_0$ is also diagonal in this basis and can be expanded as
\begin{align}
\label{W0a}
    W_{0}(\bx_1,\bk_1,\bx_2,\bk_2,t)=\sum_{i=1}^{4} a_i(\bx_1,\bk_1,\bx_2,\bk_2,t)\bv_i(\bk_1,\bk_2)\bv_i^T(\bk_1,\bk_2).
\end{align}
At order $O(\sqrt{\epsilon})$ we find that 
\begin{align} \label{Order1b}
    \nonumber&\hat{A}(\bk_1-\bK_1/2,\bk_2-\bK_2/2)\hat{W}_{1}(\bx_1,\bK_1,\bk_1,\bx_2,\bK_2,\bk_2)\\
    \nonumber&- \hat{W}_{1}(\bk_1',\bK_1,\bk_1,\bk_2',\bK_2,\bk_2)\hat{A}(\bk_1+\bK_2/2,\bk_2+\bK_2/2)\\
   \nonumber &+\coup\sqrt{\rho_0}(2\pi)^3\hat{\eta}(\bK_1)\delta(\bK_2)\left[K_1W_{0}(\bx_1,\bk_1+\bK_1/2,\bx_2,\bk_2+\bK_2)\right.\\
   \nonumber &\left.-W_{0}(\bx_1,\bk_1-\bK_1/2,\bx_2,\bk_2-\bK_2)K_1^{T}\right]\\
    \nonumber &+\coup\sqrt{\rho_0}(2\pi)^3\hat{\eta}(\bK_2)\delta(\bK_1)\left[K_2W_{0}(\bx_1,\bk_1+\bK_1,\bx_2,\bk_2+\bK_2/2)\right.\\
    &\left.-W_{0}(\bx_1,\bk_1-\bK_1,\bx_2,\bk_2-\bK_2/2)K_2^{T}\right]=0.
\end{align}
Although $W_1$ is not diagonal, we can still decompose its Fourier transform $\hat{W_1}$ as
\begin{align}
      \nonumber &\hat{W}_1(\bx_1,\bK_1,\bk_1,\bx_2,\bK_2,\bk_2)\\
      &=\sum_{i,j} w_{i,j}(\bx_1,\bK_1,\bk_1,\bx_2,\bK_2,\bk_2)\bv_{i}(\bk_1-\bK_1/2,\bk_2-\bK_2/2)\bv_{j}^{T}(\bk_1+\bK_1/2,\bk_2+\bK_2/2).
\end{align}
Multiplying (\ref{Order1b}) on the left by $\bv_{m}^{T}(\bk_1-\bK_1/2,\bk_2-\bK_2/2)$ and the right by $\bv_{n}(\bk_1+\bK_1/2,\bk_2+\bK_2/2)$, we obtain
\begin{align}
   \nonumber &(\lambda_{m}(\bk_1-\bK_1/2,\bk_2-\bK_2/2)-\lambda_{n}(\bk_1+\bK_1/2,\bk_2+\bK_2/2)+i\theta)w_{m,n}(\bx_1,\bK_1,\bk_1,\bx_2,\bK_2,\bk_2)\\
    \nonumber =& \coup\sqrt{\rho_0}(2\pi)^3\eta(\bK_1)\delta(\bK_2)\left[a_{m}(\bk_1-\bK_1/2,\bk_2-\bK_2/2)K_{1,m,n}(\bk_1-\bK_1/2,\bk_2-\bK_2/2,\bk_1+\bK_1/2,\bk_2+\bK_2/2)\right]\\
   \nonumber +&\coup\sqrt{\rho_0}(2\pi)^3\eta(\bK_2)\delta(\bK_1)\left[a_{m}(\bk_1-\bK_1/2,\bk_2-\bK_2/2)K_{2,m,n}(\bk_1-\bK_1/2,\bk_2-\bK_2/2,\bk_1+\bK_1/2,\bk_2+\bK_2/2)\right]\\
     \nonumber-& \coup\sqrt{\rho_0}(2\pi)^3\eta(\bK_1)\delta(\bK_2)\left[a_{n}(\bk_1+\bK_1/2,\bk_2+\bK_2/2)K_{1,m,n}(\bk_1+\bK_1/2,\bk_2+\bK_2/2,\bk_1-\bK_1/2,\bk_2-\bK_2/2)\right]\\
    -& \coup\sqrt{\rho_0}(2\pi)^3\eta(\bK_2)\delta(\bK_1)\left[a_{n}(\bk_1+\bK_1/2,\bk_2+\bK_2/2)K_{2,m,n}(\bk_1+\bK_1/2,\bk_2+\bK_2/2,\bk_1-\bK_1/2,\bk_2-\bK_2/2)\right]\ ,
\end{align}
where 
\begin{align}
    K_{1,m,n}(\bk_1,\bk_2,\bq_1,\bq_2)&=\bv_{m}^{T}(\bk_1,\bk_2)K_1\bv_{n}(\bq_1,\bq_2)\ , \\
    K_{2,m,n}(\bk_1,\bk_2,\bq_1,\bq_2)&=\bv_{m}^{T}(\bk_1,\bk_2)K_2\bv_{n}(\bq_1,\bq_2) \ .
\end{align}
Here $\theta\to 0^+$ is a regularizing parameter. At order $O(\epsilon)$ we find that
\begin{align} 
\label{Order2b}
    \nonumber&i\partial_t W_{0}(\bx_1,\bk_1,\bx_2,\bk_2,t)= LW_2(\bx_1,\bX_1,\bk_1,\bx_2,\bX_2,\bk_2,t)
    -\frac{i}{2} MW_{0}(\bx_1,\bk_1,\bx_2,\bk_2)
    -\frac{i}{2} W_{0}(\bx_1,\bk_1,\bx_2,\bk_2)M \\
    \nonumber
      &+\coup\sqrt{\rho_0}\int\frac{\dd^3 q}{(2\pi)^3}e^{i\bq\cdot\bX_1}\,\hat{\eta}(\bq)\left[K_1W_{1}(\bx_1,\bX_1,\bk_1+\bq/2,\bx_2,\bX_2,\bk_2)-W_{1}(\bx_1,\bX_1,\bk_1-\bq/2,\bx_2,\bX_2,\bk_2)K_1^{T}\right]\\
    &+\coup\sqrt{\rho_0}\int\frac{\dd^3 q}{(2\pi)^3}e^{i\bq\cdot\bX_2}\,\hat{\eta}(\bq)\left[K_2W_{1}
(\bx_1,\bX_1,\bk_1,\bx_2,\bX_2,\bk_2+\bq/2)-W_{1}(\bx_1,\bX_1,\bk_1,\bx_2,\bX_2,\bk_2-\bq/2)K_2^{T}\right]\, ,
\end{align}
where
\begin{align}
M = \begin{bmatrix}c\hat\bk_1\cdot\nabla_{\bx_1}+c\hat\bk_2\cdot\nabla_{\bx_2} & 0 & 0 & 0\\0 &c\hat\bk_2\cdot\nabla_{\bx_2} & 0 & 0 \\ 0 & 0 & c\hat\bk_1\cdot\nabla_{\bx_1} & 0 \\ 0 & 0 & 0 & 0\end{bmatrix}
\end{align}
and
\begin{align}
   \nonumber &LW_2(\bx_1,\bX_1,\bk_1,\bx_2,\bX_2,\bk_2,t)\\
   \nonumber&=\int\frac{\dd^3 K_1}{(2\pi)^3}\frac{\dd^3 K_2}{(2\pi)^3}e^{i\bK_1\cdot\bX_1+i\bK_2\cdot\bX_2}\hat{A}(\bk_1-\bK_1/2,\bk_2-\bK_2/2)\hat{W}_{2}(\bx_1,\bK_1,\bk_1,\bx_2,\bK_2,\bk_2,t)\\
    -&\hat{W}_{2}(\bx_1,\bK_1,\bk_1,\bx_2,\bK_2,\bk_2,t)\hat{A}(\bk_1+\bK_1/2,\bk_2+\bK_2/2).
\end{align}
In order to obtain the equations satisfied by the $a_i$, we multiply (\ref{Order2b}) on the left by $\bv_i^{T}(\bk_1,\bk_2)$ and the right by $\bv_i(\bk_1,\bk_2)$, and take the ensemble average. In order to close the hierarchy of equations, we assume that $\langle \bv_i^T LW_2 \bv_i\rangle=0$, which corresponds to the assumption that $W_2$ is statistically stationary with respect to the fast variables $\bX_1$ and $\bX_2$. Following procedures similar to those in Appendix C, we find that
\begin{align} \label{transport3}
    \nonumber \frac{1}{c}\partial_t a_{i}+&f_{1i}(\bk_1,\bk_2)\hat\bk_1\cdot\nabla_{\bx_1}a_{i}+f_{2i}(\bk_1,\bk_2)\hat\bk_2\cdot\nabla_{\bx_2}a_{i}
    \nonumber+\mu_{1i}(\bk_1,\bk_2)a_{i}(\bk_1,\bk_2)+ \mu_{2i}(\bk_1,\bk_2)a_{i}(\bk_1,\bk_2)\\
    &=\mu_{1i}(\bk_1,\bk_2)\int d^2 \hat{K}A(\hat{\bk_1},\hat{\bK},\vert\bk_1\vert)a_{i}(\vert\bk_1\vert\hat{\bK},\bk_2)+ \mu_{2i}(\bk_1,\bk_2)\int d^2 \hat{K}A(\hat{\bk_2},\hat{\bK},\vert\bk_2\vert)a_{i}(\bk_1,\vert\bk_2\vert\hat{\bK}) \ ,
\end{align}
where
\begin{align}
    f_{1i}(\bk_1,\bk_2) &= \bv_{i1}(\bk_1,\bk_2)^2+\bv_{i3}(\bk_1,\bk_2)^2\, ,\\
    f_{2i}(\bk_1,\bk_2) &= \bv_{i1}(\bk_1,\bk_2)^2+\bv_{i2}(\bk_1,\bk_2)^2\, ,\\
    \mu_{1i}(\bk_1,\bk_2) &=\frac{g^2\rho_0\pi K_{1,i,i}(\bk_1,\bk_2,\bk_1,\bk_2)^2\vert\bk_1\vert^2}{\vert\partial_{\bk_1}\lambda_i(\bk_1,\bk_2)\vert}\, , \\
    \mu_{2i}(\bk_1,\bk_2) &= \frac{g^2\rho_0\pi K_{2,i,i}(\bk_1,\bk_2,\bk_1,\bk_2)^2\vert\bk_2\vert^2}{\vert\partial_{\bk_2}\lambda_i(\bk_1,\bk_2)\vert}\, , \\
    A(\hbk_1,\hbk_2,k)&=\frac{\tilde{C}(k(\hbk_1-\hbk_2))}{\displaystyle\int d^2\hat{k'}\tilde{C}(k(\hbk_1-\hbk'))}.
\end{align}
\subsection{Diffusion approximation}
In this section we consider the diffusion limit of the kinetic equation (\ref{transport3}). We again specialize to the case of white-noise correlations, which leads to the phase function $A={1}/{(4\pi)}$, corresponding to isotropic scattering. Making use of the diffusion approximation developed in section 5.2, we see that the first angular moments
\begin{align}
\label{uiai}
u_{i}(\bx_1,\vert\bk_1\vert,\bx_2,\vert\bk_2\vert,t)=\int d\hbk_1 d\hbk_2 a_{i}(\bx_1,\bk_1,\bx_2,\bk_2,t) 
\end{align}
satisfy the equations
\begin{align} 
\label{diffusion2}
     \partial_t u_{i} - D_{1i}(\vert\bk_1\vert,\vert\bk_2\vert)\Delta_{\bx_1}u_{i} - D_{2i}(\vert\bk_1\vert,\vert\bk_2\vert)\Delta_{\bx_2}u_{i} =0\  ,
\end{align}
where
\begin{align}
    D_{1i} &= \frac{cf_{1i}^2}{3\mu_{1i}}  \ , \\
    D_{2i} &= \frac{cf_{2i}^2}{3\mu_{2i}} \ .
\end{align}

Suppose that initially there are two photons in the field localized around the points $\br_1$ and $\br_2$ in a volume of linear size $l_s$. The corresponding initial conditions are given by
\begin{align}
    \sqrt{2}\psi_2(\bx_1,\bx_2,0) &= \frac{1}{C}\left[e^{-\vert\bx_1-\br_1\vert^2/2l_s^2}e^{-\vert\bx_2-\br_2\vert^2/2l_s^2}+e^{-\vert\bx_1-\br_2\vert^2/2l_s^2}e^{-\vert\bx_2-\br_1\vert^2/2l_s^2}\right]\, ,\\
    \psi_1(\bx_1,\bx_2,0) &= a(\bx_1,\bx_2,0) = 0\, ,
\end{align}
where
\begin{align}
    C = \| e^{-\vert\bx_1-\br_1\vert^2/2l_s^2}e^{-\vert\bx_2-\br_2\vert^2/2l_s^2}+e^{-\vert\bx_1-\br_2\vert^2/2l_s^2}e^{-\vert\bx_2-\br_1\vert^2/2l_s^2} \|_{L_{\bx_1,\bx_2}^2} \ .
\end{align}
Note that this corresponds to an entangled two-photon state.

The above initial conditions imply initial conditions for the Wigner transform $W_0$. The initial conditions for the modes $a_i$ are determined by solving the linear system
\begin{align}
   \begin{bmatrix} (W_0)_{11}(\bx_1,\bk_1,\bx_2,\bk_2,0)\\ (W_0)_{22}(\bx_1,\bk_1,\bx_2,\bk_2,0)\\(W_0)_{33}(\bx_1,\bk_1,\bx_2,\bk_2,0)\\(W_0)_{44}(\bx_1,\bk_1,\bx_2,\bk_2,0) \end{bmatrix} = V(\bk_1,\bk_2)\begin{bmatrix} a_1(\bx_1,\bk_1,\bx_2,\bk_2,0)\\ a_2(\bx_1,\bk_1,\bx_2,\bk_2,0)\\  a_3(\bx_1,\bk_1,\bx_2,\bk_2,0)\\ a_4(\bx_1,\bk_1,\bx_2,\bk_2,0)\end{bmatrix} \, ,
\end{align}
where
\begin{align}
    (V(\bk_1,\bk_2))_{ij}=\bv_{ji}(\bk_1,\bk_2)^2.
\end{align}
It follows that the initial conditions for the first angular moments are given by
\begin{align}
     u_{i}(\bx_1,\vert\bk_1\vert,\bx_2,\vert\bk_2\vert,0)=\int d\hbk_1 d\hbk_2 a_{i}(\bx_1,\bk_1,\bx_2,\bk_2,0).
\end{align}
The diffusion equations (\ref{diffusion2}) can then be solved using (\ref{soln}). Combining this result with
(\ref{diag_1})--(\ref{diag_3}), (\ref{W0a}) and (\ref{uiai}) we see that the average probability densities are given by
\begin{align} 
   \nonumber &\langle\vert\psi_2(\bx_1,\bx_2,t)\vert^2\rangle\\
 \nonumber& = \frac{1}{2C^2}\sum_{i=1}^{4}\int_0^{\infty} \int_0^{\infty} dk_1 dk_2 k_1^2 k_2^2 e^{-l_s^2k_1^2 -l_s^2k_2^2}\bv_{1i}^2(k_1,k_2)(V^{-1})_{i1}(k_1,k_2)\left(\frac{l_s^2}{l_s^2+4tD_{1i}}\right)^{3/2}\\
  \nonumber &\times\left(\frac{l_s^2}{l_s^2+4tD_{2i}}\right)^{3/2}\left[ e^{-\vert\bx_1-\br_1\vert^2/(l_s^2+4tD_{1i})}e^{-\vert\bx_2-\br_2\vert^2/(l_s^2+4tD_{2i})}\right.\\
  \nonumber &+e^{-\vert\bx_1-\br_2\vert^2/(l_s^2+4tD_{1i})}e^{-\vert\bx_2-\br_1\vert^2/(l_s^2+4tD_{2i})}\\
  &\left.+2e^{-\vert \bx_1-\frac{(\br_1+\br_2)}{2}\vert^2/(l_s^2+4tD_{1i})}e^{-\vert \bx_2-\frac{(\br_1+\br_2)}{2}\vert^2/(l_s^2+4tD_{2i})}\frac{\sin(k_1\vert\br_1-\br_2\vert)}{k_1\vert\br_1-\br_2\vert}\frac{\sin(k_2\vert\br_1-\br_2\vert)}{k_2\vert\br_1-\br_2\vert}\right] \label{eq:twophotonfieldamplitude}\, ,
\end{align}
\begin{align}
      \nonumber &\langle\vert\psi_1(\bx_2,\bx_1,t)\vert^2\rangle\\
  \nonumber  &= \frac{2}{\rho_0 C^2}\sum_{i=1}^{4}\int_0^{\infty} \int_0^{\infty} dk_1 dk_2 k_1^2 k_2^2 e^{-l_s^2k_1^2 -l_s^2k_2^2}\bv_{2i}^2(k_1,k_2)(V^{-1})_{i1}(k_1,k_2)\left(\frac{l_s^2}{l_s^2+4tD_{1i}}\right)^{3/2}\\
  \nonumber &\times\left(\frac{l_s^2}{l_s^2+4tD_{2i}}\right)^{3/2}\left[ e^{-\vert\bx_1-\br_1\vert^2/(l_s^2+4tD_{1i})}e^{-\vert\bx_2-\br_2\vert^2/(l_s^2+4tD_{2i})}\right.\\
  \nonumber &+e^{-\vert\bx_1-\br_2\vert^2/(l_s^2+4tD_{1i})}e^{-\vert\bx_2-\br_1\vert^2/(l_s^2+4tD_{2i})}\\
  &\left.+2e^{-\vert \bx_1-\frac{(\br_1+\br_2)}{2}\vert^2/(l_s^2+4tD_{1i})}e^{-\vert \bx_2-\frac{(\br_1+\br_2)}{2}\vert^2/(l_s^2+4tD_{2i})}\frac{\sin(k_1\vert\br_1-\br_2\vert)}{k_1\vert\br_1-\br_2\vert}\frac{\sin(k_2\vert\br_1-\br_2\vert)}{k_2\vert\br_1-\br_2\vert}\right]\label{eq:twophotonmixedamplitute}\, ,
\end{align}
\begin{align}
      \nonumber & \langle\vert a(\bx_1,\bx_2,t)\vert^2\rangle\\
  \nonumber&= \frac{1}{2\rho_0^2 C^2}\sum_{i=1}^{4}\int_0^{\infty} \int_0^{\infty} dk_1 dk_2 k_1^2 k_2^2 e^{-l_s^2k_1^2 -l_s^2k_2^2}\bv_{4i}^2(k_1,k_2)(V^{-1})_{i1}(k_1,k_2)\left(\frac{l_s^2}{l_s^2+4tD_{1i}}\right)^{3/2}\\
  \nonumber &\times\left(\frac{l_s^2}{l_s^2+4tD_{2i}}\right)^{3/2}\left[ e^{-\vert\bx_1-\br_1\vert^2/(l_s^2+4tD_{1i})}e^{-\vert\bx_2-\br_2\vert^2/(l_s^2+4tD_{2i})}\right.\\
  \nonumber &+e^{-\vert\bx_1-\br_2\vert^2/(l_s^2+4tD_{1i})}e^{-\vert\bx_2-\br_1\vert^2/(l_s^2+4tD_{2i})}\\
  &\left.+2e^{-\vert \bx_1-\frac{(\br_1+\br_2)}{2}\vert^2/(l_s^2+4tD_{1i})}e^{-\vert \bx_2-\frac{(\br_1+\br_2)}{2}\vert^2/(l_s^2+4tD_{2i})}\frac{\sin(k_1\vert\br_1-\br_2\vert)}{k_1\vert\br_1-\br_2\vert}\frac{\sin(k_2\vert\br_1-\br_2\vert)}{k_2\vert\br_1-\br_2\vert}\right]\label{eq:twophotonatomamplitude}.
\end{align}
We note that at long times, the average probability densities decay algebraically according to 
\begin{align} 
    \langle\vert\psi_2(\bx_1,\bx_2,t)\vert^2\rangle & =\frac{B_{11}}{t^3}-\frac{B_{21}(\bx_1,\bx_2,\br_1,\br_2)}{t^4} \label{eq:twophotonfieldasymptote} \, ,\\
    \langle\vert\psi_1(\bx_1,\bx_2,t)\vert^2\rangle & =\frac{4B_{12}}{\rho_0 t^3}-\frac{4B_{22}(\bx_1,\bx_2,\br_1,\br_2)}{\rho_0 t^4}\label{eq:twophotonmixedasymptote}\, ,\\
    \langle\vert a(\bx_1,\bx_2,t)\vert^2\rangle & =\frac{B_{14}}{\rho_0^2 t^3}-\frac{B_{24}(\bx_1,\bx_2,\br_1,\br_2)}{\rho_0^2 t^4}\label{eq:twophotonatomasymptote}\ ,
\end{align}
where the constants $B_{1j}$ and $B_{2j}$, $j=1,2,4$ are given by
\begin{align}
    B_{1j} & = \frac{1}{C^2}\sum_{i=1}^{4}\int_0^{\infty} \int_0^{\infty} dk_1 dk_2 k_1^2 k_2^2 e^{-l_s^2k_1^2 -l_s^2k_2^2}\bv_{ji}^2(k_1,k_2)(V^{-1})_{i1}(k_1,k_2)\left(\frac{l_s^2}{4D_{1i}}\right)^{3/2}\\
  \nonumber &\times\left(\frac{l_s^2}{4D_{2i}}\right)^{3/2}\left[1+1\frac{\sin(k_1\vert\br_1-\br_2\vert)}{k_1\vert\br_1-\br_2\vert}\frac{\sin(k_2\vert\br_1-\br_2\vert)}{k_2\vert\br_1-\br_2\vert}\right]\, ,\\
   B_{2j} & = \frac{1}{2C^2}\sum_{i=1}^{4}\int_0^{\infty} \int_0^{\infty} dk_1 dk_2 k_1^2 k_2^2 e^{-l_s^2k_1^2 -l_s^2k_2^2}\bv_{ji}^2(k_1,k_2)(V^{-1})_{i1}(k_1,k_2)\left(\frac{l_s^2}{4D_{1i}}\right)^{3/2}\\
  \nonumber &\times\left(\frac{l_s^2}{4D_{2i}}\right)^{3/2}\left[ \frac{\vert\bx_1-\br_1\vert^2}{4D_{1i}}+\frac{\vert\bx_2-\br_2\vert^2}{4D_{2i}}+\frac{\vert\bx_1-\br_2\vert^2}{4D_{1i}}+\frac{\vert\bx_2-\br_1\vert^2}{4D_{2i}}\right.\\
  &\left.+2\left(\frac{\vert \bx_1-\frac{(\br_1+\br_2)}{2}\vert^2}{4D_{1i}}+\frac{\vert \bx_2-\frac{(\br_1+\br_2)}{2}\vert^2}{4D_{2i}}\right)\frac{\sin(k_1\vert\br_1-\br_2\vert)}{k_1\vert\br_1-\br_2\vert}\frac{\sin(k_2\vert\br_1-\br_2\vert)}{k_2\vert\br_1-\br_2\vert}\right].
\end{align}

\begin{figure}[t]
    \centering
    \includegraphics{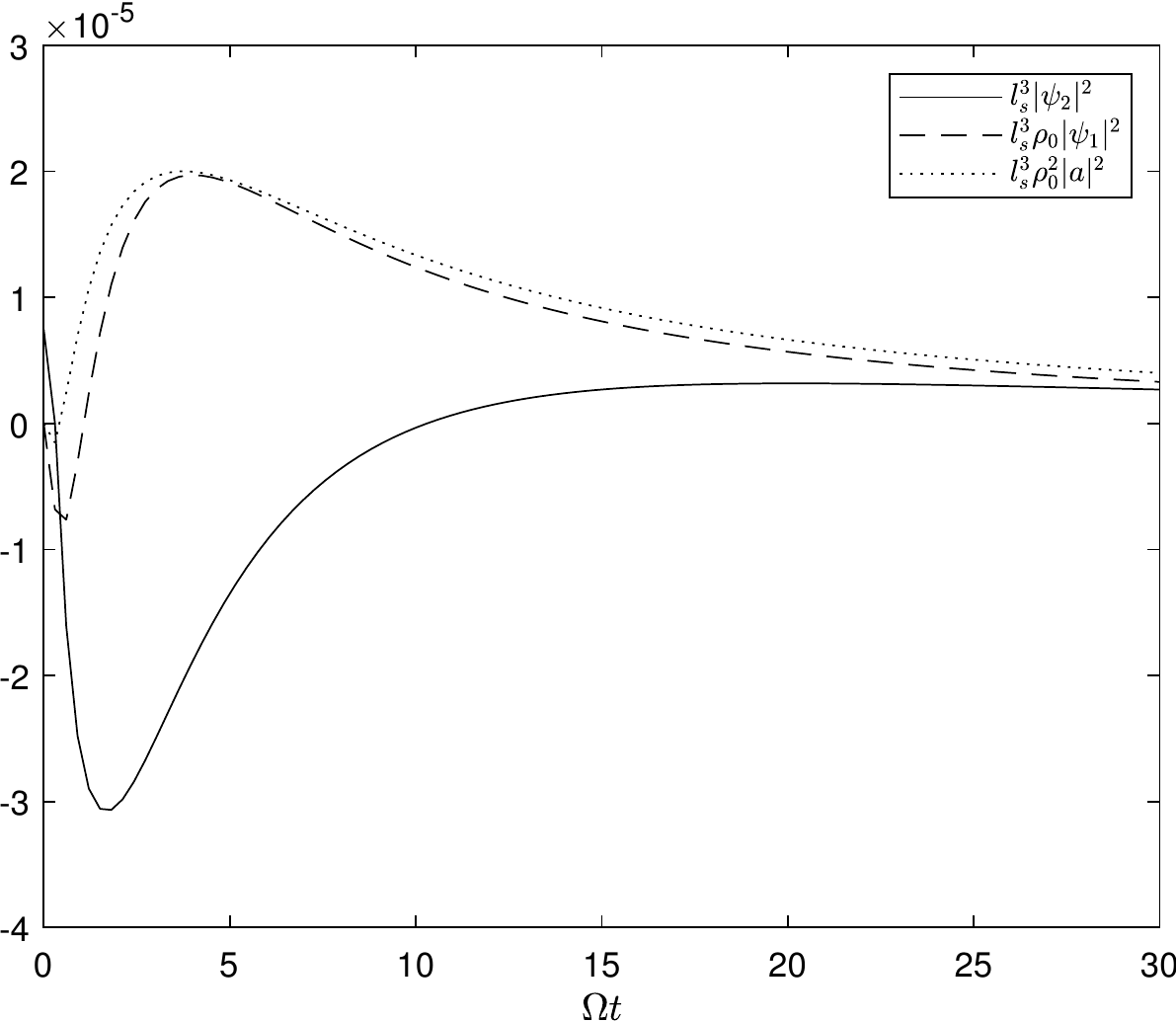}
    \caption{Atomic, one-photon, and two-photon probability densities in a random medium.}
    \label{fig:amplitudes}
\end{figure}

\begin{figure}[t]
    \centering
    \includegraphics{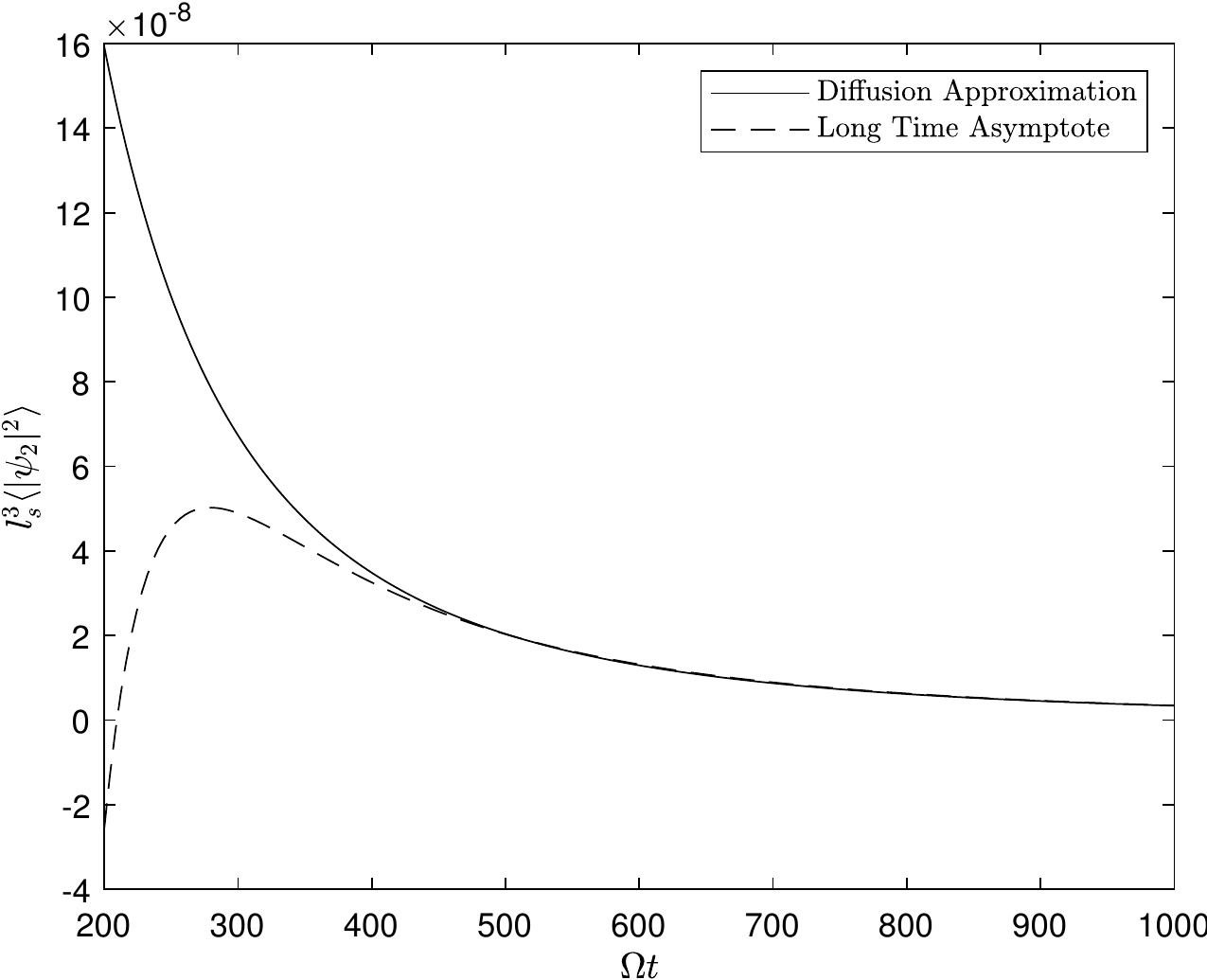}
    \caption{Comparison of diffusion approximation and long-time asymptote for $\langle\vert\psi_2\vert^2\rangle$.}
    \label{fig:twophotonfieldasymptote}
\end{figure}

\begin{figure}[t]
    \centering
    \includegraphics{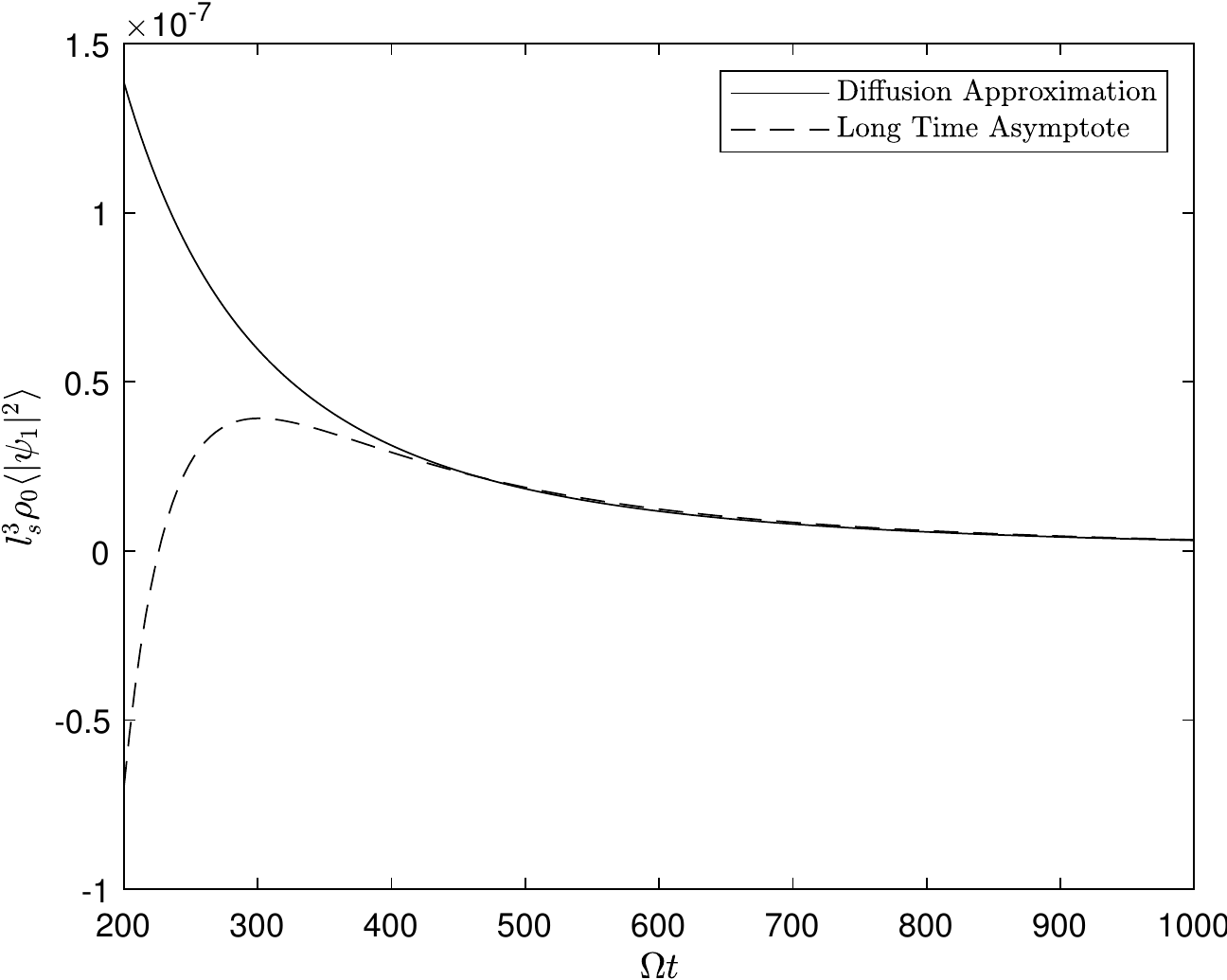}
    \caption{Comparison of diffusion approximation and long-time asymptote for $\langle\vert\psi_1\vert^2\rangle$.}
    \label{fig:twophotonmixedasymptote}
\end{figure}

\begin{figure}[t]
    \centering
    \includegraphics{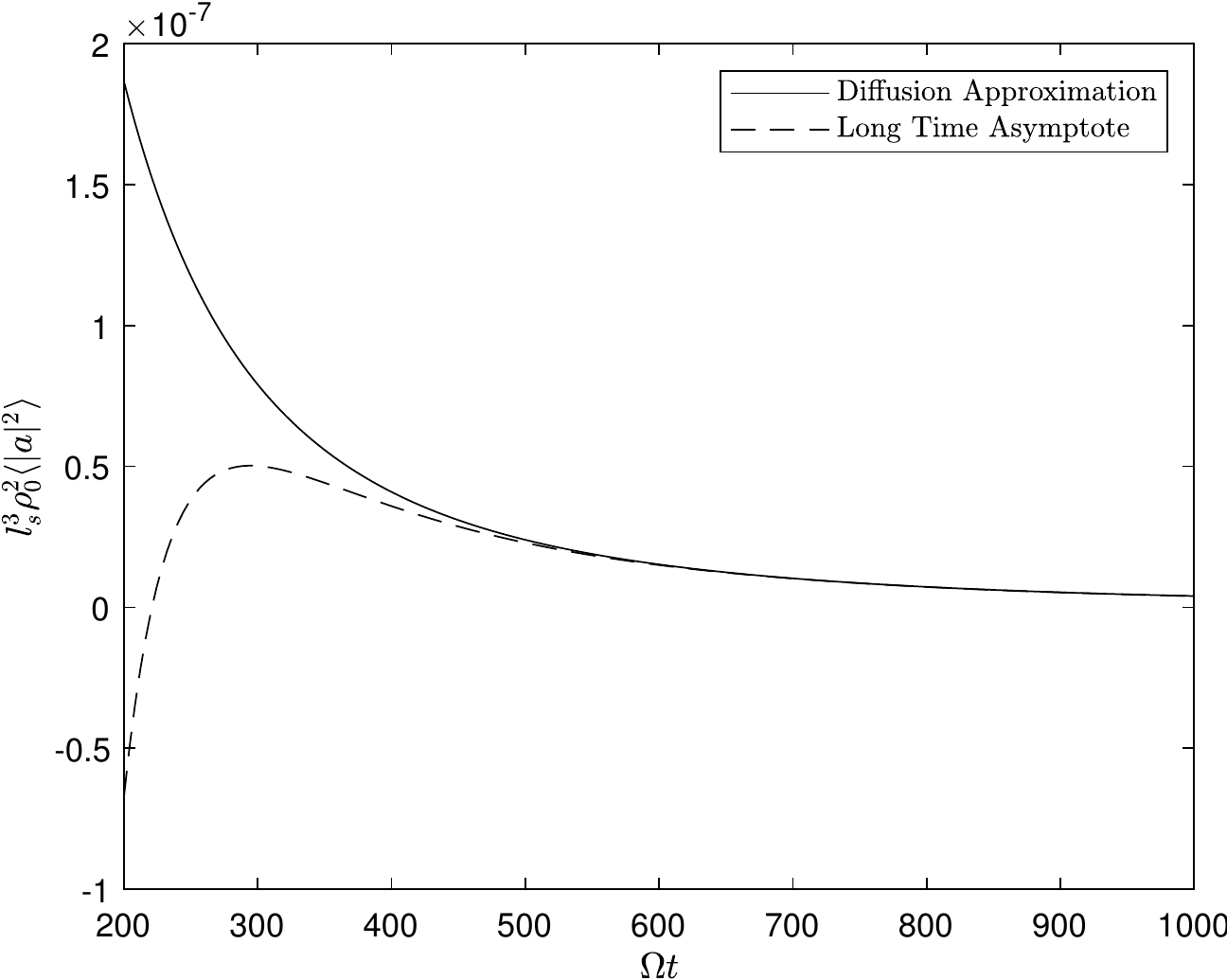}
    \caption{Comparison of diffusion approximation and long-time asymptote for $\langle\vert a \vert^2\rangle$. }
    \label{fig:twophotonatomasymptote}
\end{figure}

In order to illustrate the above results, we consider the case of isotropic scattering and set the dimensionless quantities ${\Omega}/{\sqrt{\rho_0}\coup}={c}/{l_s\sqrt{\rho_0}\coup}=1$. In addition, we choose  $\bx_1=(l_s,0,0)$, $\bx_2=(-l_s,0,0)$, $\br_1=(0,0,l_s)$ and $\br_2=(0,0,-l_s)$, so that the distances from the points of excitation ($\br_1$ and $\br_2$) to the points of detection are equal to $l_s$. In Figure~\ref{fig:amplitudes} we plot the time dependence of  the probability densities $a$, $|\langle \psi_1 \rangle |^2$ and $|\langle \psi_2 \rangle |^2$. We note that the negative values of these quantities are due to the breakdown of the diffusion approximation at short times.
We observe that the two-photon probability density increases before eventually decaying.
 A comparison of these results with the asymptotic formulas (\ref{eq:twophotonfieldasymptote}),  (\ref{eq:twophotonmixedasymptote}) and (\ref{eq:twophotonatomasymptote}) is shown in Figures \ref{fig:twophotonfieldasymptote}, \ref{fig:twophotonmixedasymptote} and \ref{fig:twophotonatomasymptote}. There is good agreement at long times.
 
 \section{Discussion}
We have investigated the propagation of two-photon light in a random medium of two-level atoms.
Our main results are kinetic equations that govern the behavior of the field and atomic probability densities. Several topics for further research are apparent. An alternative derivation of these equations may be possible using diagrammatic perturbation theory rather than multiscale asymptotic analysis. This is the case for the classical theory of wave propagation in random media, where a comparative exposition of the two approaches has been presented in~\cite{Caze_2015}. It would be of some interest to quantify the evolution of the entanglement of a two-photon state as it propagates through the medium. One approach to this problem is to introduce a suitable measure of entanglement such as the entropy defined by the singular values of the two-photon amplitude~\cite{Schotland_2016}. Here the evolution of the entanglement of an initially entangled state is of particular importance, especially in applications to communications and imaging. Finally, accounting for atomic motion is a challenging problem. One possible approach is to view the density $\rho$ as a quantum field whose dynamics must be taken into account.


\appendix
\section{Derivation of the System (\ref{eq:dynamics3})}
Here we derive the system (\ref{eq:dynamics3}). We begin by computing both sides of the time-dependent Schrodinger equation  (\ref{eq:schr}), employing the Hamiltonian $H$ and the state (\ref{state}). The left-hand side is equal to
\begin{align}
 \nonumber   i\hbar\partial_t\vert\Psi\rangle = \int d^3 x_1 d^3 x_2 & \left(i\hbar\partial_t\psi_2(\bx_1,\bx_2,t)\phi^{\dagger}(\bx_1)\phi^{\dagger}(\bx_2)
+ i\hbar\partial_t\psi_1(\bx_1,\bx_2,t)\rho(\bx_1)\sigma^{\dagger}(\bx_1)\phi^{\dagger}(\bx_2)\right.\\
+&\left. i\hbar\partial_ta(\bx_1,\bx_2,t)\rho(\bx_1)\rho(\bx_2)\sigma^{\dagger}(\bx_1)\sigma^{\dagger}(\bx_2)\right)\vert 0\rangle \ ,
\end{align}
while the right-hand side is given by
\begin{align}
   \nonumber H\vert\Psi\rangle & =\hbar\int d^3x d^3 x_1 d^3 x_2\left\{c(-\Delta)^{1/2}\phi^{\dagger}(\bx)\phi(\bx) +\Omega\rho(\bx)\sigma^{\dagger}(\bx)\sigma(\bx) + \coup\rho(\bx)\left(\phi^{\dagger}(\bx)\sigma(\bx)+\phi(\bx)\sigma^{\dagger}(\bx) \right)\right\} \\
   \nonumber &\times\{\psi_2(\bx_1,\bx_2,t)\phi^{\dagger}(\bx_1)\phi^{\dagger}(\bx_2)
+ \psi_1(\bx_1,\bx_2,t)\rho(\bx_1)\sigma^{\dagger}(\bx_1)\phi^{\dagger}(\bx_2) \\
\nonumber&+ a(\bx_1,\bx_2,t)\rho(\bx_1)\rho(\bx_2)\sigma^{\dagger}(\bx_1)\sigma^{\dagger}(\bx_2)\}\vert 0\rangle\\
\nonumber&=\hbar\int d^3x d^3 x_1 d^3 x_2 \{ \psi_2(\bx_1,\bx_2,t)c(-\Delta)^{1/2}\phi^{\dagger}(\bx)\phi(\bx)\phi^{\dagger}(\bx_1)\phi^{\dagger}(\bx_2)\\
\nonumber&+\psi_1(\bx_1,\bx_2,t)\rho(\bx_1)c(-\Delta)^{1/2}\phi^{\dagger}(\bx)\phi(\bx)\sigma^{\dagger}(\bx_1)\phi^{\dagger}(\bx_2)\\
\nonumber&+\psi_1(\bx_1,\bx_2,t)\rho(\bx_1)\Omega\rho(\bx)\sigma^{\dagger}(\bx)\sigma(\bx)\sigma^{\dagger}(\bx_1)\phi^{\dagger}(\bx_2)\\
\nonumber&+a(\bx_1,\bx_2,t)\rho(\bx_1)\rho(\bx_2)\Omega\rho(\bx)\sigma^{\dagger}(\bx)\sigma(\bx)\sigma^{\dagger}(\bx_1)\sigma^{\dagger}(\bx_2)\\
\nonumber\nonumber&+\psi_1(\bx_1,\bx_2,t)\rho(\bx_1)\coup\rho(\bx)\phi^{\dagger}(\bx)\sigma(\bx)\sigma^{\dagger}(\bx_1)\phi^{\dagger}(\bx_2)\\
\nonumber&+a(\bx_1,\bx_2,t)\rho(\bx_1)\rho(\bx_2)\coup\rho(\bx)\phi^{\dagger}(\bx)\sigma(\bx)\sigma^{\dagger}(\bx_1)\sigma^{\dagger}(\bx_2)\\
\nonumber&+\psi_1(\bx_1,\bx_2,t)\rho(\bx_1)\coup\rho(\bx)\phi(\bx)\sigma^{\dagger}(\bx)\sigma^{\dagger}(\bx_1)\phi^{\dagger}(\bx_2)\\
\nonumber&+\psi_2(\bx_1,\bx_2,t)\coup\rho(\bx)\phi(\bx)\sigma^{\dagger}(\bx)\phi^{\dagger}(\bx_1)\phi^{\dagger}(\bx_2)\}\vert 0\rangle.
\end{align}
Using the commutation and anticommutation relations (\ref{commutation}) and (\ref{anticommutation}) we arrive at
\begin{align}
\nonumber&\hbar\int d^3 x_1 d^3 x_2 \{ (c(-\Delta_{\bx_1})^{1/2}\psi_2(\bx_1,\bx_2,t)+c(-\Delta_{\bx_2})^{1/2}\psi_2(\bx_1,\bx_2,t))\phi^{\dagger}(\bx_1)\phi^{\dagger}(\bx_2)\\
\nonumber&+c(-\Delta_{\bx_2})^{1/2}\psi_1(\bx_1,\bx_2,t)\rho(\bx_1)\sigma^{\dagger}(\bx_1)\phi^{\dagger}(\bx_2)+\psi_1(\bx_1,\bx_2,t)\rho(\bx_1)\Omega\sigma^{\dagger}(\bx_1)\phi^{\dagger}(\bx_2)\\
\nonumber&+2 a(\bx_1,\bx_2,t)\rho(\bx_1)\rho(\bx_2)\Omega\sigma^{\dagger}(\bx_1)\sigma^{\dagger}(\bx_2)+\psi_1(\bx_1,\bx_2,t)\rho(\bx_1)\coup\phi^{\dagger}(\bx_1)\phi^{\dagger}(\bx_2)\\
\nonumber&-2\coup a(\bx_1,\bx_2,t)\rho(\bx_1)\rho(\bx_2)\phi^{\dagger}(\bx_2)\sigma^{\dagger}(\bx_1)+
\coup\psi_1(\bx_1,\bx_2,t)\rho(\bx_1)\sigma^{\dagger}(\bx_1)\sigma^{\dagger}(\bx_2)\\
&+2\coup\psi_2(\bx_1,\bx_2,t)\rho(\bx_1)\sigma^{\dagger}(\bx_1)\phi^{\dagger}(\bx_2)\}\vert 0\rangle.
\end{align}
Computing the inner products 
\begin{align}
   \langle 0\vert\phi(\bx_1)\phi(\bx_2)i\hbar\partial_t\vert\Psi\rangle &= 2i\hbar\partial_t\psi_2(\bx_1,\bx_2,t) \ ,\\
    \nonumber\langle 0\vert\phi(\bx_1)\phi(\bx_2)H\vert\Psi\rangle &= 2(c(-\Delta_{\bx_1})^{1/2}+c(-\Delta_{\bx_2})^{1/2})\psi_2(\bx_1,\bx_2,t)\\
    &+\coup\rho(\bx_1)\psi_1(\bx_1,\bx_2,t)+\coup\rho(\bx_2)\psi_1(\bx_2,\bx_1,t)\, , 
\end{align}
yields (\ref{eq:dynamics1}). The inner products
\begin{align}
     \langle 0\vert\phi(\bx_2)\sigma(\bx_1)\rho(\bx_1)i\hbar\partial_t\vert\Psi\rangle &= \rho(\bx_1)i\hbar\partial_t\psi_1(\bx_1,\bx_2,t) \ , \\
    \nonumber\langle 0\vert\phi(\bx_2)\sigma(\bx_1)\rho(\bx_1)H\vert\Psi\rangle &= (c(-\Delta_{\bx_2})^{1/2}+\Omega)\rho(\bx_1)\psi_1(\bx_1,\bx_2,t)\\
    &+2\coup\rho(\bx_1)\psi_2(\bx_1,\bx_2,t)-2\coup\rho(\bx_1)\rho(\bx_2)a(\bx_1,\bx_2,t)\, , 
\end{align}
give (\ref{eq:dynamics2}), while
\begin{align}
    \langle 0\vert\sigma(\bx_1)\sigma(\bx_2)\rho(\bx_1)\rho(\bx_2)i\hbar\partial_t\vert\Psi\rangle &= 2\rho(\bx_1)\rho(\bx_2)i\hbar\partial_t a(\bx_1,\bx_2,t) \ , \\
    \nonumber\langle 0\vert\sigma(\bx_1)\sigma(\bx_2)\rho(\bx_1)\rho(\bx_2)H\vert\Psi\rangle &= \coup\rho(\bx_1)\rho(\bx_2)\psi_1(\bx_1,\bx_2,t)-\coup\rho(\bx_1)\rho(\bx_2)\psi_1(\bx_2,\bx_1,t)\\
    &+4\Omega\rho(\bx_1)\rho(\bx_2)a(\bx_1,\bx_2,t)\  ,
\end{align}
gives (\ref{eq:dynamics3}).
\section{Derivation of the Liouville Equation (\ref{Liouville1})}
Here we derive ~(\ref{Liouville1}). We first define 
\begin{align}
     \Phi_{\epsilon}(\bx_1,\bx_1',\bx_2,\bx_2',t)=\bPsi_{\epsilon}(\bx_1-\epsilon\bx_1'/2,\bx_2-\epsilon\bx_2'/2,t)\bPsi_{\epsilon}^\dagger(\bx_1+\epsilon\bx_1'/2,\bx_2 +\epsilon\bx_2'/2,t) \ .
\end{align}
The Wigner transform is defined by
\begin{align}
    W_{\epsilon}(\bx_1,\bk_1,\bx_2,\bk_2,t) =\int\frac{d^3 x_1'}{(2\pi)^3}\frac{d^3 x_2'}{(2\pi)^3}e^{-i\bk_1\cdot\bx_1'-i\bk_2\cdot\bx_2'} \Phi_{\epsilon}(\bx_1,\bx_1',\bx_2,\bx_2',t) \ .
\end{align}
We begin by computing $i\epsilon\partial_t W_{\epsilon}$:
\begin{align}
&\nonumber 
i\epsilon\partial_t W_{\epsilon} =\int\frac{d^3 x_1'}{(2\pi)^3}\frac{d^3 x_2'}{(2\pi)^3}e^{-i\bk_1\cdot\bx_1'-i\bk_2\cdot\bx_2'} i\epsilon\partial_t\Phi_{\epsilon}(\bx_1,\bx_1',\bx_2,\bx_2',t)\\
    &\nonumber=\int\frac{d^3 x_1'}{(2\pi)^3}\frac{d^3 x_2'}{(2\pi)^3}e^{-i\bk_1\cdot\bx_1'-i\bk_2\cdot\bx_2'}\left[\left\{A_{\epsilon}(\bx_1-\epsilon\bx_1'/2,\bx_2-\epsilon\bx_2'/2) \Phi_{\epsilon}(\bx_1,\bx_1',\bx_2,\bx_2',t)\right.\right.\\
    \nonumber&\left.+ \sqrt{\epsilon}\coup\sqrt{\frac{\rho_0}{2}}(\eta(\bx_1/\epsilon-\bx_1'/2)+\eta(\bx_2/\epsilon-\bx_2'/2))K\ \Phi_{\epsilon}(\bx_1,\bx_1',\bx_2,\bx_2',t)\right\}\\
    &\nonumber-  \Phi_{\epsilon}(\bx_1,\bx_1',\bx_2,\bx_2',t)A_{\epsilon}(\bx_1+\epsilon\bx_1'/2,\bx_2 +\epsilon\bx_2'/2) \\
    \nonumber& + \left.\left.\sqrt{\epsilon}\coup\sqrt{\frac{\rho_0}{2}}(\eta(\bx_1/\epsilon+\bx_1'/2)+\eta(\bx_2/\epsilon+\bx_2'/2)) \Phi_{\epsilon}(\bx_1,\bx_1',\bx_2,\bx_2',t)K^{T}\right\}\right]\\
    \nonumber&=\int\frac{d^3 x_1'}{(2\pi)^3}\frac{d^3 x_2'}{(2\pi)^3}e^{-i\bk_1\cdot\bx_1'-i\bk_2\cdot\bx_2'}\left\{A_{\epsilon}(\bx_1-\epsilon\bx_1'/2,\bx_2-\epsilon\bx_2'/2)\Phi_{\epsilon}(\bx_1,\bx_1',\bx_2,\bx_2',t) \right.\\
    \nonumber&\left.-\Phi_{\epsilon}(\bx_1,\bx_1',\bx_2,\bx_2',t)A_{\epsilon}(\bx_1+\epsilon\bx_1'/2,\bx_2 +\epsilon\bx_2'/2)\right\}\\
    \nonumber&+\sqrt{\epsilon}\coup\sqrt{\frac{\rho_0}{2}}\int\frac{d^3 x_1'}{(2\pi)^3}\frac{d^3 x_2'}{(2\pi)^3}e^{-i\bk_1\cdot\bx_1'-i\bk_2\cdot\bx_2'}\left\{\eta(\bx_1/\epsilon-\bx_1'/2)K\Phi_{\epsilon}(\bx_1,\bx_1',\bx_2,\bx_2',t)\right.\\
    \nonumber&\left.-\eta(\bx_1/\epsilon+\bx_1'/2)\Phi_{\epsilon}(\bx_1,\bx_1',\bx_2,\bx_2',t)K^{T}\right\}\\
     \nonumber&+\sqrt{\epsilon}\coup\sqrt{\frac{\rho_0}{2}}\int\frac{d^3 x_1'}{(2\pi)^3}\frac{d^3 x_2'}{(2\pi)^3}e^{-i\bk_1\cdot\bx_1'-i\bk_2\cdot\bx_2'}\left\{\eta(\bx_2/\epsilon-\bx_2'/2)K\Phi_{\epsilon}(\bx_1,\bx_1',\bx_2,\bx_2',t)\right.\\
    &\left.-\eta(\bx_2/\epsilon+\bx_2'/2)\Phi_{\epsilon}(\bx_1,\bx_1',\bx_2,\bx_2',t)K^{T}\right\}.
\end{align}
The first term becomes
\begin{align}
    \nonumber&\int\frac{d^3 x_1'}{(2\pi)^3}\frac{d^3 x_2'}{(2\pi)^3}e^{-i\bk_1\cdot\bx_1'-i\bk_2\cdot\bx_2'}\left\{A_{\epsilon}(\bx_1-\epsilon\bx_1'/2,\bx_2-\epsilon\bx_2'/2)\Phi_{\epsilon}(\bx_1,\bx_1',\bx_2,\bx_2',t) \right.\\
    \nonumber&\left.-\Phi_{\epsilon}(\bx_1,\bx_1',\bx_2,\bx_2',t)A_{\epsilon}(\bx_1+\epsilon\bx_1'/2,\bx_2 +\epsilon\bx_2'/2)\right\}\\
    \nonumber&=\int\frac{\dd^3 k_1'}{(2\pi)^3}\frac{\dd^3 k_2'}{(2\pi)^3}e^{i\bx_1\cdot\bk_1'+i\bx_2\cdot\bk_2'}\left\{\hat{A}(\bk_1-\epsilon\bk_1'/2,\bk_2-\epsilon\bk_2'/2)\hat{W}_{\epsilon}(\bk_1',\bk_1,\bk_2',\bk_2)\right.\\
    &-\left. \hat{W}_{\epsilon}(\bk_1',\bk_1,\bk_2',\bk_2)\hat{A}(\bk_1+\epsilon\bk_1'/2,\bk_2+\epsilon\bk_2'/2)\right\}\, ,
\end{align}
where 
\begin{align}
    (A_{\epsilon}f)(\bx_1,\bx_2)=\int\frac{d^3 k_1}{(2\pi)^3}\frac{d^3 k_2}{(2\pi)^3}e^{i\bx_1\cdot\bk_1+i\bx_2\cdot\bk_2}\hat{A}(\epsilon\bk_1,
    \epsilon\bk_2)\hat{f}(\bk_1,\bk_2) \ .
\end{align}
Likewise, the second term becomes
\begin{align}
    \nonumber&\sqrt{\epsilon}\coup\sqrt{\frac{\rho_0}{2}}\int\frac{d^3 x_1'}{(2\pi)^3}\frac{d^3 x_2'}{(2\pi)^3}e^{-i\bk_1\cdot\bx_1'-i\bk_2\cdot\bx_2'}\left\{\eta(\bx_1/\epsilon-\bx_1'/2)K\Phi_{\epsilon}(\bx_1,\bx_1',\bx_2,\bx_2',t)\right.\\
    \nonumber&\left.-\eta(\bx_1/\epsilon+\bx_1'/2)\Phi_{\epsilon}(\bx_1,\bx_1',\bx_2,\bx_2',t)K^{T}\right\}\\
    \nonumber&=\sqrt{\epsilon}\coup\sqrt{\frac{\rho_0}{2}}\int\frac{d^3 x_1'}{(2\pi)^3}\frac{d^3 x_2'}{(2\pi)^3}\frac{d^3 q}{(2\pi)^3}e^{-i\bk_1\cdot\bx_1'-i\bk_2\cdot\bx_2'+i\bq\cdot(\bx_1/\epsilon-\bx_1'/2)}\hat{\eta}(\bq)K\Phi_{\epsilon}(\bx_1,\bx_1',\bx_2,\bx_2',t)\\
    \nonumber&-\sqrt{\epsilon}\coup\sqrt{\frac{\rho_0}{2}}\int\frac{d^3 x_1'}{(2\pi)^3}\frac{d^3 x_2'}{(2\pi)^3}\frac{d^3 q}{(2\pi)^3}e^{-i\bk_1\cdot\bx_1'-i\bk_2\cdot\bx_2'+i\bq\cdot(\bx_1/\epsilon+\bx_1'/2)}\hat{\eta}(\bq)\Phi_{\epsilon}(\bx_1,\bx_1',\bx_2,\bx_2',t)K^{T}\\
    \nonumber&=\sqrt{\epsilon}\coup\sqrt{\frac{\rho_0}{2}}\int\frac{d^3 q}{(2\pi)^3}e^{i\bq\cdot\bx_1/\epsilon}\hat{\eta}(\bq)K W_{\epsilon}(\bx_1,\bk_1+\bq/2,\bx_2,\bk_2,t)\\
    \nonumber&-\sqrt{\epsilon}\coup\sqrt{\frac{\rho_0}{2}}\int\frac{d^3 q}{(2\pi)^3}e^{i\bq\cdot\bx_1/\epsilon}\hat{\eta}(\bq)W_{\epsilon}(\bx_1,\bk_1-\bq/2,\bx_2,\bk_2,t)K^{T}\\
    &=\sqrt{\epsilon}\coup\sqrt{\frac{\rho_0}{2}}\int\frac{d^3 q}{(2\pi)^3}e^{i\bq\cdot\bx_1/\epsilon}\hat{\eta}(\bq)\left[K W_{\epsilon}(\bx_1,\bk_1+\bq/2,\bx_2,\bk_2,t)-W_{\epsilon}(\bx_1,\bk_1-\bq/2,\bx_2,\bk_2,t)K^{T}\right].
\end{align}
Finally, the third term becomes
\begin{align}
    \nonumber&\sqrt{\epsilon}\coup\sqrt{\frac{\rho_0}{2}}\int\frac{d^3 x_1'}{(2\pi)^3}\frac{d^3 x_2'}{(2\pi)^3}e^{-i\bk_1\cdot\bx_1'-i\bk_2\cdot\bx_2'}\left\{\eta(\bx_2/\epsilon-\bx_2'/2)K\Phi_{\epsilon}(\bx_1,\bx_1',\bx_2,\bx_2',t)\right.\\
    \nonumber&\left.-\eta(\bx_2/\epsilon+\bx_2'/2)\Phi_{\epsilon}(\bx_1,\bx_1',\bx_2,\bx_2',t)K^{T}\right\}\\
    &=\sqrt{\epsilon}\coup\sqrt{\frac{\rho_0}{2}}\int\frac{d^3 q}{(2\pi)^3}e^{i\bq\cdot\bx_1/\epsilon}\hat{\eta}(\bq)\left[K W_{\epsilon}(\bx_1,\bk_1,\bx_2,\bk_2+\bq/2,t)-W_{\epsilon}(\bx_1,\bk_1,\bx_2,\bk_2-\bq/2,t)K^{T}\right].
\end{align}
\section{Derivation of the Kinetic Equation (\ref{transport1})}
Here we derive the kinetic equation (\ref{transport1}) which is satisfied by the quantity $a_{+}$. 
The first two terms are elementary and so we must compute 
\begin{align}
\langle\bv_+^T L_1 W_1 \bv_+\rangle \ , \quad \langle\bv_+^T L_2 W_1 \bv_+\rangle \ .
\end{align}
We compute each of the above in two steps. The first term is
\begin{align}
    \nonumber&\langle \int\frac{\dd^3 q}{(2\pi)^3}e^{i\bq\cdot\bX_1}\,\hat{\eta}(\bq)\left[\bv_{+}^{T}(\bk_1,\bk_2)KW_{1}(\bx_1,\bX_1,\bk_1+\bq/2,\bx_2,\bX_2,\bk_2)\bv_{+}(\bk_1,\bk_2)\right]\rangle \\
    \nonumber=&\langle\int\frac{\dd^3 q}{(2\pi)^3}\frac{\dd^3 K_1}{(2\pi)^3}\frac{\dd^3 K_2}{(2\pi)^3}e^{i\bq\cdot\bX_1+i\bK_1\cdot\bX_1+i\bK_2\cdot\bX_2}\,\hat{\eta}(\bq)\\
    \nonumber&\times\left[\bv_{+}^{T}(\bk_1,\bk_2)K\hat{W}_{1}(\bx_1,\bK_1,\bk_1+\bq/2,\bx_2,\bK_2,\bk_2)\bv_{+}(\bk_1,\bk_2)\right]\rangle \\
   \nonumber=&\langle\int\frac{\dd^3 q}{(2\pi)^3}\frac{\dd^3 K_1}{(2\pi)^3}\frac{\dd^3 K_2}{(2\pi)^3}e^{i\bq\cdot\bX_1+i\bK_1\cdot\bX_1+i\bK_2\cdot\bX_2}\,\hat{\eta}(\bq) \bv_{+}^{T}(\bk_1,\bk_2)K\\
    \nonumber&\times\left\{\sum_{m,n=\pm} w_{m,n}(\bK_1,\bk_1+\bq/2,\bK_2,\bk_2)\bv_{m}(\bk_1+\bq/2-\bK_1/2,\bk_2-\bK_2/2)\right.\\
    \nonumber&\left.\times\bv_{n}^{T}(\bk_1+\bq/2+\bK_1/2,\bk_2+\bK_2/2)\bv_{+}(\bk_1,\bk_2)\right\}\rangle.
\end{align}
Next we substitute the expression $w_{m,n}$ using equation (\ref{eq:wmn}) to arrive at 
\begin{align}
    \nonumber&\langle\int\frac{\dd^3 q}{(2\pi)^3}\frac{\dd^3 K_1}{(2\pi)^3}\frac{\dd^3 K_2}{(2\pi)^3}e^{i\bq\cdot\bX_1+i\bK_1\cdot\bX_1+i\bK_2\cdot\bX_2}\,\hat{\eta}(\bq) \bv_{+}^{T}(\bk_1,\bk_2)K\\
    \nonumber&\times\left\{\sum_{m,n=\pm} g\sqrt{\frac{\rho_0}{2}}(2\pi)^3\left\{\eta(\bK_1)\delta(\bK_2)+\eta(\bK_2)\delta(\bK_1)\right\}\right\}\\
    \nonumber&\times\left(\frac{\left[a_{m}(\bk_1+\bq/2-\bK_1/2,\bk_2-\bK_2/2)K_{m,n}(\bk_1+\bq/2-\bK_1/2,\bk_2-\bK_2/2,\bk_1+\bq/2+\bK_1/2,\bk_2+\bK_2/2)\right]}{\lambda_{m}(\bk_1+\bq/2-\bK_1/2,\bk_2-\bK_2/2)-\lambda_{n}(\bk_1+\bq/2+\bK_1/2,\bk_2+\bK_2/2)+i\theta}\right.\\
   \nonumber &-\left.\frac{\left[a_{n}(\bk_1+\bq/2+\bK_1/2,\bk_2+\bK_2/2)K_{m,n}(\bk_1+\bq/2+\bK_1/2,\bk_2+\bK_2/2,\bk_1+\bq/2-\bK_1/2,\bk_2-\bK_2/2)\right]}{\lambda_{m}(\bk_1+\bq/2-\bK_1/2,\bk_2-\bK_2/2)-\lambda_{n}(\bk_1+\bq/2+\bK_1/2,\bk_2+\bK_2/2)+i\theta}\right)\\
    \nonumber&\times\bv_{m}(\bk_1+\bq/2-\bK_1/2,\bk_2-\bK_2/2)\bv_{n}^{T}(\bk_1+\bq/2+\bK_1/2,\bk_2+\bK_2/2)\bv_{+}(\bk_1,\bk_2)\rangle.
\end{align}
We separate the above into two terms and use the orthogonality of the basis $\{\bv_i\}$ to see that $n=+$ in the first term. 
We thus obtain
\begin{align}
    \nonumber&=g\sqrt{\frac{\rho_0}{2}} \int\frac{\dd^3 K_1}{(2\pi)^3}\,\hat{C}(\bK_1)\sum_{m=\pm} \left(\frac{\left[a_{m}(\bk_1-\bK_1,\bk_2)K_{m,+}(\bk_1-\bK_1,\bk_2,\bk_1,\bk_2)\right]}{\lambda_{m}(\bk_1-\bK_1,\bk_2)-\lambda_{+}(\bk_1,\bk_2)+i\theta}\right.\\
    \nonumber&-\left.\frac{\left[a_{+}(\bk_1,\bk_2)K_{m,+}(\bk_1,\bk_2,\bk_1-\bK_1,\bk_2)\right]}{\lambda_{m}(\bk_1-\bK_1,\bk_2)-\lambda_{+}(\bk_1,\bk_2)+i\theta}\right)K_{+,m}(\bk_1,\bk_2,\bk_1-\bK_1,\bk_2)\\
    \nonumber&+g\sqrt{\frac{\rho_0}{2}}\int\frac{\dd^3 K_2}{(2\pi)^3}\,e^{i\bK_2\cdot(\bX_2-\bX_1)}\hat{C}(\bK_2) \bv_{+}^{T}(\bk_1,\bk_2)K\sum_{m,n=\pm}\\
    \nonumber&\times\left(\frac{\left[a_{m}(\bk_1-\bK_2/2,\bk_2-\bK_2/2)K_{m,n}(\bk_1-\bK_2/2,\bk_2-\bK_2/2,\bk_1-\bK_2/2,\bk_2+\bK_2/2)\right]}{\lambda_{m}(\bk_1-\bK_2/2,\bk_2-\bK_2/2)-\lambda_{n}(\bk_1-\bK_2/2,\bk_2+\bK_2/2)+i\theta}\right.\\
   \nonumber &-\left.\frac{\left[a_{n}(\bk_1-\bK_2/2,\bk_2+\bK_2/2)K_{m,n}(\bk_1-\bK_2/2,\bk_2+\bK_2/2,\bk_1-\bK_2/2,\bk_2-\bK_2/2)\right]}{\lambda_{m}(\bk_1-\bK_2/2,\bk_2-\bK_2/2)-\lambda_{n}(\bk_1-\bK_2/2,\bk_2+\bK_2/2)+i\theta}\right)\\
    &\times\bv_{m}(\bk_1-\bK_2/2,\bk_2-\bK_2/2)\bv_{n}^{T}(\bk_1-\bK_2/2,\bk_2+\bK_2/2)\bv_{+}(\bk_1,\bk_2).
\end{align}
As $\epsilon\to 0$ the above converges to
\begin{align}
    &g\sqrt{\frac{\rho_0}{2}} \int\frac{\dd^3 K}{(2\pi)^3}\,\hat{C}(\bk_1-\bK)\sum_{m=\pm} \left(\frac{\left[a_{m}(\bK,\bk_2)K_{m,+}(\bK,\bk_2,\bk_1,\bk_2)\right]}{\lambda_{m}(\bK,\bk_2)-\lambda_{+}(\bk_1,\bk_2)+i\theta}\right.\\
    &-\left.\frac{\left[a_{+}(\bk_1,\bk_2)K_{m,+}(\bk_1,\bk_2,\bK,\bk_2)\right]}{\lambda_{m}(\bK,\bk_2)-\lambda_{+}(\bk_1,\bk_2)+i\theta}\right)K_{+,m}(\bk_1,\bk_2,\bK,\bk_2)\ ,
\end{align}
which follows from  the Riemann-Lebesgue Lemma and the fact that $W_0$ is independent of the fast variables $\bX_1$ and $\bX_2$. Similarly the other three terms become
\begin{align}
   \nonumber&-g\sqrt{\frac{\rho_0}{2}}\int\frac{\dd^3 K}{(2\pi)^3}\,\hat{C}(\bk_1-\bK) \sum_{n=\pm} 
    \left(\frac{\left[a_{+}(\bk_1,\bk_2)K_{+,n}(\bk_1,\bk_2,\bK,\bk_2)\right]}{\lambda_{+}(\bk_1,\bk_2)-\lambda_{n}(\bK,\bk_2)+i\theta}
    \right.\\
   \nonumber &-\left.\frac{\left[a_{n}(\bK,\bk_2)K_{+,n}(\bK,\bk_2,\bk_1,\bk_2)\right]}{\lambda_{+}(\bk_1,\bk_2)-\lambda_{n}(\bK,\bk_2)+i\theta}\right)K_{+,n}(\bk_1,\bk_2,\bK,\bk_2)\\
   \nonumber &+g\sqrt{\frac{\rho_0}{2}}\int\frac{\dd^3 K}{(2\pi)^3}\,\hat{C}(\bk_2-\bK)\sum_{m=\pm} 
    \left(\frac{\left[a_{m}(\bk_1,\bK)K_{m,+}(\bk_1,\bK,\bk_1,\bk_2)\right]}{\lambda_{m}(\bk_1,\bK)-\lambda_{+}(\bk_1,\bk_2)+i\theta}\right.\\
   \nonumber &-\left.\frac{\left[a_{+}(\bk_1,\bk_2)K_{m,+}(\bk_1,\bk_2,\bk_1,\bK)\right]}{\lambda_{m}(\bk_1,\bK)-\lambda_{+}(\bk_1,\bk_2)+i\theta}\right)K_{+,m}(\bk_1,\bk_2,\bk_1,\bK)\\
   \nonumber &-g\sqrt{\frac{\rho_0}{2}}\int\frac{\dd^3 K}{(2\pi)^3}\,\hat{C}(\bk_2-\bK) \sum_{+,n=\pm} \left(\frac{\left[a_{+}(\bk_1,\bk_2)K_{+,n}(\bk_1,\bk_2,\bk_1,\bK)\right]}{\lambda_{+}(\bk_1,\bk_2)-\lambda_{n}(\bk_1,\bK)+i\theta}\right.\\
   &-\left.\frac{\left[a_{n}(\bk_1,\bK)K_{+,n}(\bk_1,\bK,\bk_1,\bk_2)\right]}{\lambda_{+}(\bk_1,\bk_2)-\lambda_{n}(\bk_1,\bK)+i\theta}\right)K_{+,n}(\bk_1,\bk_2,\bk_1,\bK).
\end{align}
Making use of the identity
\begin{align}
      \lim_{\theta\to 0}\left(\frac{1}{x-i\theta}-\frac{1}{x+i\theta}\right)= 2\pi i\delta(x)\ ,
\end{align}
we see that we must have $m,n=+$, to ensure that the support of the delta function is nonempty since $\lambda_+$ and $\lambda_-$ are never equal. Thus the above equation simplifies as
\begin{align}
    \nonumber&-i\pi \coup\sqrt{\frac{\rho_0}{2}} \int\frac{\dd^3 K}{(2\pi)^3}\,\hat{C}(\bk_1-\bK)\delta(\lambda_{+}(\bK,\bk_2)-\lambda_{+}(\bk_1,\bk_2))\\
     \nonumber&\times\left(a_{+}(\bk_1,\bk_2)K_{+,+}(\bk_1,\bk_2,\bK,\bk_2)^2 - a_{+}(\bK,\bk_2)K_{+,+}(\bK,\bk_2,\bk_1,\bk_2)K_{+,+}(\bk_1,\bk_2,\bK,\bk_2)\right)\\
    \nonumber &-i\pi \coup\sqrt{\frac{\rho_0}{2}} \int\frac{\dd^3 K}{(2\pi)^3}\,\hat{C}(\bk_2-\bK)\delta(\lambda_{+}(\bk_1,\bK)-\lambda_{+}(\bk_1,\bk_2))\\
    &\times\left(a_{+}(\bk_1,\bk_2)K_{+,+}(\bk_1,\bk_2,\bk_1,\bK)^2 - a_{+}(\bk_1,\bK)K_{+,+}(\bk_1,\bK,\bk_1,\bk_2)K_{+,+}(\bk_1,\bk_2,\bk_1,\bK)\right).
\end{align}
Putting everything together, we see that $a_{+}$ satisfies the equation
\begin{align} \label{eq:c1}
     &\nonumber\frac{1}{c}\partial_t a_{+}+\left[\frac{(\lambda_{+}(\bk_1,\bk_2)-d(\bk_1,\bk_2)-\Omega)^2+\coup^2\rho_0}{(\lambda_{+}(\bk_1,\bk_2)-d(\bk_1,\bk_2)-\Omega)^2+2\coup^2\rho_0}\right]\left(\hat\bk_1\cdot\nabla_{\bx_1}+\hat\bk_2\cdot\nabla_{\bx_2}\right)a_{+}\\
    =&-\pi \coup\frac{\rho_0}{2} \int\frac{\dd^3 K}{(2\pi)^3}\,\hat{C}(\bk_1-\bK)\delta(\lambda_{+}(\bK,\bk_2)-\lambda_{+}(\bk_1,\bk_2))\\
     \nonumber&\times\left(a_{+}(\bk_1,\bk_2)K_{+,+}(\bk_1,\bk_2,\bK,\bk_2)^2 - a_{+}(\bK,\bk_2)K_{+,+}(\bK,\bk_2,\bk_1,\bk_2)K_{+,+}(\bk_1,\bk_2,\bK,\bk_2)\right)\\
    \nonumber &-\pi \coup^2\frac{\rho_0}{2} \int\frac{\dd^3 K}{(2\pi)^3}\,\hat{C}(\bk_2-\bK)\delta(\lambda_{+}(\bk_1,\bK)-\lambda_{+}(\bk_1,\bk_2))\\
    &\times\left(a_{+}(\bk_1,\bk_2)K_{+,+}(\bk_1,\bk_2,\bk_1,\bK)^2 - a_{+}(\bk_1,\bK)K_{+,+}(\bk_1,\bK,\bk_1,\bk_2)K_{+,+}(\bk_1,\bk_2,\bk_1,\bK)\right).
\end{align}
The delta function can be simplified using the identity
\begin{align}
    \delta(g(x))=\frac{\delta(x-x_0)}{\vert g'(x_0)\vert}
\end{align}
where $g$ has a single real root at $x=x_0$. Thus
\begin{align}
     \delta(\lambda_{+}(\bK,\bk_2)-\lambda_{+}(\bk_1,\bk_2))&=\frac{\delta(\vert\bK\vert-\vert\bk_1\vert)}{\frac{c}{4}\left\vert{c(\vert\bk\vert+\vert\bk_2\vert)}/{2}-\Omega+\frac{3}{2}\sqrt{({c(\vert\bk\vert+\vert\bk_2\vert)}/{2}-\Omega)^2+8g^2\rho_0}\right\vert}\ .
\end{align}
Hence ~(\ref{eq:c1}) becomes
\begin{align}
    \nonumber&\frac{1}{c}\partial_t a_{+}+\left[\frac{(\lambda_{+}(\bk_1,\bk_2)-d(\bk_1,\bk_2)-\Omega)^2+\coup^2\rho_0}{(\lambda_{+}(\bk_1,\bk_2)-d(\bk_1,\bk_2)-\Omega)^2+2\coup^2\rho_0}\right]\left(\hat\bk_1\cdot\nabla_{\bx_1}+\hat\bk_2\cdot\nabla_{\bx_2}\right)a_{+}\\
    \nonumber=&-\frac{2g^2\rho_0\pi}{c}\frac{K_{+,+}(\bk_1,\bk_2,\bk_1,\bk_2)^2}{\left\vert d(\bk_1,\bk_2)-\Omega+\frac{3}{2}\sqrt{(d(\bk_1,\bk_2)-\Omega)^2+8g^2\rho_0}\right\vert}\vert\bk_1\vert^2 \\
     \nonumber&\times\int\frac{d\hbk'}{(2\pi)^3}\,\hat{C}(\vert\bk_1\vert(\hbk_1-\hbk'))\left(a_{+}(\bk_1,\bk_2) - a_{+}(\vert\bk_1\vert\hbk',\bk_2)\right)\\
    \nonumber &-\frac{2g^2\rho_0\pi}{c}\frac{K_{+,+}(\bk_1,\bk_2,\bk_1,\bk_2)^2}{\left\vert d(\bk_1,\bk_2)-\Omega+\frac{3}{2}\sqrt{(d(\bk_1,\bk_2)-\Omega)^2+8g^2\rho_0}\right\vert}\vert\bk_2\vert^2 \\
    &\times \int\frac{d\hbk'}{(2\pi)^3}\,\hat{C}(\vert\bk_2\vert(\hbk_2-\hbk'))\left(a_{+}(\bk_1,\bk_2)- a_{+}(\bk_1,\vert\bk_2\vert\hbk')\right)\ ,
\end{align}
as desired. 

\section{Bosonic Model}
Here we consider the bosonic analog of the model described in section 2. The physical implications of the bosonic model will be explored elsewhere.

We suppose that the atomic field $\sigma$ satisfies the bosonic commutation relations
\begin{align}
    [\sigma(\bx),\sigma^{\dagger}(\bx')] &= \frac{1}{\rho(\bx)}\delta(\bx-\bx')\ , \\
    [\sigma(\bx),\sigma(\bx')] &= 0 \ .
\end{align}
The Hamiltonian (\ref{Htot}) remains unchanged. As in section 2, we assume that the state $\vert\Psi\rangle$ is of the form 
\begin{align}
\vert\Psi\rangle = \int d^3 x_1 d^3 x_2 & \left(\psi_2(\bx_1,\bx_2,t)\phi^{\dagger}(\bx_1)\phi^{\dagger}(\bx_2)
+ \psi_1(\bx_1,\bx_2,t)\rho(\bx_1)\sigma^{\dagger}(\bx_1)\phi^{\dagger}(\bx_2)\right.\\
+\nonumber&\left. a(\bx_1,\bx_2,t)\rho(\bx_1)\rho(\bx_2)\sigma^{\dagger}(\bx_1)\sigma^{\dagger}(\bx_2)\right)\vert 0\rangle \ .
\end{align}
The amplitudes $\psi_2$ and $a$ are now both taken to be symmetric,
\begin{equation}
\psi_2(\bx_2,\bx_1,t) = \psi_2(\bx_1,\bx_2,t) \ ,  \quad a(\bx_2,\bx_1,t) = a(\bx_1,\bx_2,t) \ ,
\end{equation}
consistent with the bosonic  character of the fields. Making use of the Schrodinger equation (\ref{eq:schr}), we find that $\psi_1$, $\psi_2$, and $a$ satisfy the system of equations
\begin{align}
\label{eq:Bdynamics1}
&\nonumber i\partial_t\psi_2(\bx_1,\bx_2,t) =c(-\Delta_{\bx_1})^{1/2}\psi_2(\bx_1,\bx_2,t)+c(-\Delta_{\bx_2})^{1/2}\psi_2(\bx_1,\bx_2,t) \\
&+\frac{g}{2}(\rho(\bx_1)\psi_1(\bx_1,\bx_2,t)+\rho(\bx_2)\psi_1(\bx_2,\bx_1,t)) \ , \\
 & \nonumber\rho(\bx_1)i\partial_t \psi_1(\bx_1,\bx_2,t) = 2\coup\rho(\bx_1)\psi_2(\bx_1,\bx_2,t)+\rho(\bx_1)\left[c(-\Delta_{\bx_2})^{1/2}+\Omega\right]\psi_1(\bx_1,\bx_2,t)\\
\label{eq:Bdynamics2}
 &+2\coup\rho(\bx_1)\rho(\bx_2)a(\bx_1,\bx_2,t) \ , \\
  &  \rho(\bx_1)\rho(\bx_2)i\partial_t a(\bx_1,\bx_2,t) =\rho(\bx_1)\rho(\bx_2)\frac{g}{2}(\psi_1(\bx_2,\bx_1,t)+\psi_1(\bx_1,\bx_2,t)) + 2\Omega\rho(\bx_1)\rho(\bx_2)\, a(\bx_1,\bx_2,t) \ .
\label{eq:Bdynamics3}
\end{align}
Note that there is a difference in the equations of motion for the bosonic and fermionic cases. Namely, a single term in each of (\ref{eq:dynamics2}), (\ref{eq:dynamics3}) and (\ref{eq:Bdynamics2}), (\ref{eq:Bdynamics3}) changes sign. We also note that there is no corresponding change in the single excitation theory of \cite{Kraisler_2022}.   

The bosonic model allows more than one atomic excitation to be present at the same point in space. To address this difficulty, the Hamiltonian is modified by adding an interaction term of the form
\begin{align}
H = \hbar U\int d^3 x \rho(\bx)n(\bx) \left(n(\bx)-1 \right) \ ,
\end{align}
where the number operator $n(\bx)= \sigma^{\dagger}(\bx)\sigma(\bx)$ and the repulsion $U$ is a large positive number. The above term penalizes the creation of an excitation at a point where another is present. A similar term arises in the Hubbard model for fermionic systems.


\section*{Acknowledgments}
This work was performed when the authors were members of the Department of Mathematics at University of Michigan. We thank Nick Read for valuable discussions. This work was supported in part by the NSF grant DMS-1912821 and the AFOSR grant FA9550-19-1-0320.
J.E.K. was supported, in part, by Simons Foundation Math + X Investigator Award No. 376319 (Michael I. Weinstein).

\section*{Author Declarations}

\subsection*{Conflict of Interest}

The authors have no conflicts to disclose.

\section*{Data Sharing}

Data sharing is not applicable to this article as no new data were created or analyzed in this study.

\end{document}